\documentstyle[12pt]{article}
\makeatletter
\renewcommand{\theequation}{\thesection.\arabic{equation}}
\@addtoreset{equation}{section}
\makeatother

\begin{document}

\begin{flushright}
DFPD10/TH/03\\
February, 2010\\
\end{flushright}
\vspace{20pt}

\pagestyle{empty}
\baselineskip15pt

\begin{center}
{\large\bf Pure Spinor Approach to Type IIA Superstring Sigma Models and Free
Differential Algebras }
{\vskip 1mm }

\vspace{20mm}

Mario Tonin
          \footnote{
          E-mail address:\ mario.tonin@pd.infn.it}\\
\vspace{5mm}
          Dipartimento di Fisica, Universita degli Studi di Padova,\\
          INFN, Sezione di Padova,\\
          Via F. Marzolo 8, 35131 Padova, Italy\\

\end{center}


\vspace{5mm}
\begin{abstract}
 This paper considers the Free Differential Algebra and rheonomic
parametrization of type IIA Supergravity, extended to include the BRS
differential and the ghosts. We consider
not only the ghosts $\lambda$'s of supersymmetry  but also
the ghosts corresponding to gauge and Lorentz transformations. In this way
we can derive not only the BRS transformations of fields and ghosts but also
the standard pure spinor constraints on $\lambda$'s. Moreover the formalism
allows to derive the action for the pure spinor formulation of type IIA
superstrings in a general background, recovering the action first obtained
by Berkovits and Howe.

\end{abstract}

\newpage
\pagestyle{plain}
\pagenumbering{arabic}

\section{Introduction}
The pure spinor formulation of superstrings \cite{Ber1} is a very powerful
method to provide a covariant quantization of superstring theories in flat
superspace \cite{Ber2} - \cite{Ber5}  or in special backgrounds like for
instance $AdS_5 \times S_5$ \cite{Ber6}, but it is also
important  to describe superstrings in general backgrounds \cite{BH}-
\cite{tor1}, especially in presence of RR fluxes.

The pure spinor formulation of $\sigma$-models to  lowest order in
$\alpha'$ has been developed some time ago
in \cite{BH}, both for  heterotic and  type II superstrings (for
heterotic
strings see also \cite{Oda1}; for  first order
 $\alpha'$ corrections in the heterotic case,  at the cohomological level,
 see \cite{Cha2}.)
The method of \cite{BH} starts by writing the more general action
invariant under worldsheet conformal transformations and then derives the
constraints of the background superfields by requiring nilpotence of the
BRST charge and  holomorphicity of the BRST currents (or equivalently
invariance of the action under the BRST charge \cite{Cha1}).

Recently D'Auria, Grassi, Fre' and Trigiante \cite{tor1}, working in
the case of type IIA  \footnote{A reason to deal with the type IIA
case is the recent interest on type IIA superstings in the $AdS_4 \times
CP^3$ background \cite{dima} - \cite{Bonel}. } superstring,
 have proposed an
alternative method that reverses this procedure: they start from the
geometrical formulation of type IIA supergravity and by generalizing a
procedure \cite{Stora}, \cite{ Bon1}  well known for
Yang-Mills theories, to which we shall refer as the method of the Extended
(Free) Differential Algebra,
they derive the constraints for the ghosts and the BRST invariant
$\sigma$-model
action. However the constraints  that they obtain are
not standard. Moreover  these constraints, even if  imply for the ghosts
the same number of d.o.f.'s of the standard constraints,  as they argue, are
quite involved and lead to a complicated action.

In this paper we present a variant of the approach of \cite{tor1} for type IIA
$\sigma$-models  that instead
leads to the standard constraints on pure spinors and to the pure spinor
action of \cite{BH}. The type IIB case presumably can be treated in a similar
way and hopefully will be presented elsewhere.

The paper is organized as follows. Section II deals with the
geometrical description of type IIA supergravity. In particular,
after fixing  notation,  we  present the supergravity constraints to
be used in the next sections. For that, we  adopt  the
parametrization of the curvatures derived in \cite{tor1},
\cite{tor2}. In Section III we explain the approach of the Extended
Differential Algebra and we derive the BRST algebra and the standard
constraints on pure spinors. In Section IV we obtain the pure spinor
action. The derivation of a property needed to assure the
consistency of this action is given in the Appendix.
\section{Review of the geometrical description of D=10, type IIA
Supergravity }

The geometrical formulation of D=10, type IIA SUGRA involves the following
objects:  \\
the vector-like and
spinor-like supervielbeins $E^{a} = dZ^{M}E_{M}{}^{a}(Z)$ , $E_{L}^{\alpha}
 = dZ^{M}E_{L{} M}{}^{\alpha}(Z)$ ,   $E_{R}^{\hat \alpha} = dZ^{M}
E_{R{} M}{}^{\hat \alpha}(Z) $  which are one-superforms, the
Lorentz--group one--superform connection $\Omega^{ab}(Z)$; the NS-NS
two-superform $B_{2}(Z)$; the dilaton superfield $\phi(Z)$; the R-R
one-superform $C_{1}(Z)$ and 3-superform and $C_{3}(Z)$. The indices
$a=0,1,...9$, $\alpha = 1, ...16$ and $\hat\alpha = 1,...,16)$,are
respectively the tangent space vector and spinor indices of $D=10$
type IIA superspace. The spinor indices are those of the
Weyl-Majorana spinors with opposite chirality. $Z^{M} =
(X^{m},\theta_{L}^{\mu},\theta_{R}^{\hat\mu})$ are the superspace
coordinates\footnote{ Letters from the middle of the alphabet denote
curved indices, those from the begin of the alphabet denote tangent
space, Lorentz indices. Round brackets (square brackets) will denote
symmetrization (antisymmetrization) with $ V^{[A}V^{B]}= {1 \over
2}( V^{A}V^{B}- V^{B}V^{A})$ and $ V^{(A}V^{B)}= {1 \over 2}(
V^{A}V^{B}+ V^{B}V^{A})$. }.

Sometimes we shall use the notation that an upper (lower) index
$\hat \alpha$ is written as a lower (upper) index $\alpha$  so that,
for instance,
$$E_{R}^{\hat\alpha}
\equiv E_{R \alpha}.$$  We shall also write
$ E^{\underline{\alpha}} = ( E_{L}^{\alpha}, E_{R}^{\hat\alpha} )$
and $E^{A} = (E^{a}, E^{\underline{\alpha}})$ .

The curvatures of the superforms defined above and of the dilaton are:\\
the torsions
\begin{eqnarray}
T^{a} = \Delta E^{a} \equiv  dE^{a} - E^{b}\Omega_{b}{}^{a}
\label{2.1}
\end{eqnarray}
\begin{eqnarray}
T_{L}^{\alpha} = \Delta E_{L}^{\alpha} \equiv d  E_{L}{}^{\alpha} - E_{L}{}^{
\beta}\Omega_{\beta}{}^{\alpha}
 \label{2.2}
\end{eqnarray}
\begin{eqnarray}
T_{R}^{\hat\alpha} = \Delta E_{R}^{\hat\alpha} \equiv d  E_{R}{}^{\hat \alpha}
 - E_{R}{}^{\hat \beta}\Omega_{\hat \beta}{}^{\hat \alpha}
 \label{2.3}
\end{eqnarray}
the Lorentz curvature
\begin{eqnarray}
R^{ab} = d \Omega^{ab} - \Omega^{a}{}_{c}\Omega^{cb}
\label{2.4}
\end{eqnarray}
the NS-NS curvature
\begin{eqnarray}
H_{3} = dB_{2},
\label{2.5}
\end{eqnarray}
 the R-R curvatures
\begin{eqnarray}
G_{2} = dC_{1}
\label{2.6}
\end{eqnarray}
\begin{eqnarray}
G_{4} = d  C_{3} + B_{2} d C_{1}.
\label{2.7}
\end{eqnarray}
and
\begin{eqnarray}
F_{1} = d \phi .
\label{2.8}
\end{eqnarray}
 $\Delta $ denotes the Lorentz covariant differential and $ \Omega_{\underline{
\beta}}{}^{\underline{\alpha}} = {1 \over 4}\Omega^{ab}(\Gamma_{ab})_{
\underline{\beta}}{}^{\underline{\alpha}} \equiv (\Omega_{\beta}{}^{\alpha},
\Omega_{\hat \beta}{}^{\hat \alpha})$. The torsions and curvatures
(\ref{2.1})-(\ref{2.7})  satisfy the free differential algebra of Bianchi
identities
\begin{eqnarray}
\Delta T^{a} = - E^{b}R_{b}{}^{a}
\label{2.9}
\end{eqnarray}
\begin{eqnarray}
\Delta T^{\underline{\alpha}} = - E^{\underline{\beta}}R_{\underline{\beta}}{}^
{\underline{\alpha}}
\label{2.10}
\end{eqnarray}
where $$  R_{\underline {\alpha}}{}^{\underline{\beta}} ={1 \over 4} R^{ab}
(\Gamma_{ab})_{\underline{\alpha}}{}^{\underline{\beta}} \equiv (R_{L \alpha}
{}^{\beta}, R_{R \hat \alpha}{}^{\hat \beta}).$$
Moreover
\begin{eqnarray}
\Delta R^{ab} = 0
\label{2.11}
\end{eqnarray}
\begin{eqnarray}
dH_{3} = 0
\label{2.12}
\end{eqnarray}
\begin{eqnarray}
dG_{2} = 0
\label{2.012}
\end{eqnarray}
\begin{eqnarray}
dG_{4} = H_{3} G_{2}
\label{2.13}
\end{eqnarray}

(\ref{2.11}) is automatically satisfied if (\ref{2.9}), (\ref{2.10}) hold
(Dragon theorem \cite{Drag}).

The field equations are obtained from the Bianchi identities after
imposing suitable constraints on the torsions and curvatures
\cite{Howe},
\cite{Carr}.
There is some freedom in the choice of the constraints, related to
possible field redefinitions of supervielbeins and gauge superforms.
We will adopt the parametrization of torsions and curvatures given
in \cite{tor1}, \cite{tor2} that, in our notation, results in the
following solution of the supergravity constraints (rheonomic
parametrization):
\begin{eqnarray}
T^{a} = {1 \over 2} (E_{L}\Gamma^{a}E_{L}) +{1 \over 2} (E_{R}\Gamma^{a}E_{R})
\label{2.14}
\end{eqnarray}
\begin{eqnarray}
 T_{L}{}^{\alpha} =  {1 \over 4} [E_{L}{}^{\alpha}(E_{L}D_{R}\phi) +
{1 \over 2} (\Gamma^{ab})^{\alpha}{}_{\beta}E_{L}{}^{\beta}(D_{R}\phi
\Gamma_{ab}E_{L})] - {3 \over 4} E^{c}(\Gamma^{ab})^{\alpha}
{}_{\beta}E_{L}{}^{\beta} H_{cab}
\nonumber
\end{eqnarray}
\begin{eqnarray}
 + E^{c}(M\Gamma_{c}E_{R})^{\alpha}+ E^{b}E^{c}
T_{L{} b c}{}^{\alpha}
\label{2.15}
\end{eqnarray}
\begin{eqnarray}
 T_{R}^{\hat\alpha} =  {1 \over 4} [E_{R}{}^{\hat\alpha}(E_{R}D_{L}\phi) +
{1 \over 2} (\Gamma^{ab})^{\hat \alpha}{}_{\hat \beta}E_{R}{}^{\hat \beta}
(D_{L}\phi\Gamma_{ab}E_{R})] + {3 \over 4} E^{c}(\Gamma^{ab})^{\hat \alpha}
{}_{\hat \beta}E_{R}{}^{\hat \beta}H_{cab}
\nonumber
\end{eqnarray}
 \begin{eqnarray}
- E^{c}(E_{L}
\Gamma_{c}M )^{\hat \alpha}+ E^{b}E^{c}T_{R{} b c}{}^{\hat \alpha}
\label{2.16}
\end{eqnarray}
$$ R^{ab} = {3 \over 2} [(E_{L}\Gamma_{c}E_{L}) - (E_{R}\Gamma_{c}E_{R})]
H^{cab} + 2 (E_{L}\Gamma^{[a}M\Gamma^{b]}E_{R})$$
\begin{eqnarray}
+ E^{c}((E_{L}\Theta_{R{} c}{}^{a b}) +
(E_{R}\Theta_{L{} c}{}^{a b})) + E^{c}E^{d}R_{cd}{}^{a b}
\label{2.17}
\end{eqnarray}
\begin{eqnarray}
H_{3} = - E^{a }[(E_{L}\Gamma_{a}E_{L}) - (E_{R}\Gamma_{a}E_{R})] +
E^{a}E^{b}E^{c}H_{abc}
\label{2.18}
\end{eqnarray}
\begin{eqnarray}
G_{2} = - e^{\phi}(E_{R}E_{L}) - e^{\phi}E^{a}[(E_{L}\Gamma_{a}D_{L}\phi) -
 (E_{R}\Gamma_{a}D_{R}\phi)] + E^{a}E^{b}G_{a b}
\label{2.19}
\end{eqnarray}
$$ G_{4} =  e^{\phi}E^{a}E^{b}(E_{L}\Gamma_{ab}E_{R}) + {1 \over 3} e^{\phi}
E^{a}E^{b}E^{c}( (D_{L}\phi\Gamma_{abc}E_{L}) $$
\begin{eqnarray}
+ (D_{R}\phi\Gamma_{abc}E_{R})) + E^{a}E^{b}E^{c}E^{d}G_{abcd}
\label{2.20}
\end{eqnarray}
\begin{eqnarray}
F_{1} \equiv d\phi = E_{L}^{\alpha}D_{R \alpha}\phi + E_{R}^{\hat \alpha}
D_{L \hat \alpha}\phi  + E^{a}D_{a}\phi
\label{2.21}
\end{eqnarray}
where $$ \Theta_{R/L{} c|ab}= -((\Gamma_{a}T_{R/L{} b c}) + (\Gamma_{b}
T_{R/L{} c a})
- (\Gamma_{c}T_{R/L{} a b})) $$
and $ M^{\alpha \hat \beta}\equiv M^{\alpha}{}_{\beta}$ is a bispinor defined
as
\begin{eqnarray}
 M \equiv e^{\phi}P  = e^{\phi}[ P_{abcd}\Gamma^{abcd}
+ P_{ab}\Gamma^{ab}]
\label{2.22}
\end{eqnarray}
 with $$ P_{abcd} = - {1 \over 16} [e^{-\phi}G_{abcd} + {1 \over {12}} (D_{L}
\phi\Gamma_{abcd}D_{R}\phi)]$$ $$ P_{ab} = - {1 \over 8} [e^{-\phi}G_{ab} +
{1 \over 2} (D_{L}\phi\Gamma_{ab}D_{R}\phi)]$$
 $D_{\underline{\alpha}} \equiv (D_{R \alpha}, D_{L \hat \beta})$ and $D_{a}$
are the tangent space components of the differential $ d $.

For further purposes it is convenient to display the (0,2) sectors of eqs.
(\ref{2.15}), (\ref{2.16}) (i.e. the sectors that involve products
of two spinor-like superviewlbeins) by writing
\begin{eqnarray}
E_{L}{}^{\beta}E_{L}{}^{\gamma}X_{\beta \gamma}{}^{\alpha} \equiv
{1 \over 4} [E_{L}{}^{\alpha}(E_{L}D_{R}\phi) +
{1 \over 2} (\Gamma^{ab})^{\alpha}{}_{\beta}E_{L}{}^{\beta}(D_{R}\phi
\Gamma_{ab}E_{L})]
\nonumber
\end{eqnarray}
\begin{eqnarray}
E_{R}{}^{\hat \beta}E_{R} {}^{\hat \gamma}X_{\hat \beta \hat \gamma}{}^{
\hat \alpha} \equiv  {1 \over 4} [E_{R}{}^{\hat\alpha}(E_{R}D_{L}\phi) +
{1 \over 2} (\Gamma^{ab})^{\hat \alpha}{}_{\hat \beta}E_{R}{}^{\hat \beta}
(D_{L}\phi\Gamma_{ab}E_{R})]
\label{2.25}
\end{eqnarray}
and noting, using the Fierz identities, that the contributions of these
sectors can be rewritten as
\begin{eqnarray}
E_{L}{}^{\beta}E_{L} {}^{\gamma}X_{\beta \gamma}{}^{\alpha} \equiv
 - E_{L}{}^{\alpha}(E_{L}D_{R}\phi) +
{1 \over 2} (E_{L}\Gamma^{a}E_{L})(\Gamma_{a}D_{R}\phi)^{\alpha}
\nonumber
\end{eqnarray}
\begin{eqnarray}
E_{R}{}^{\hat \beta}E_{R} {}^{\hat \gamma}X_{\hat \beta \hat \gamma}{}^{
\hat \alpha} \equiv  - E_{R}{}^{\hat\alpha}(E_{R}D_{L}\phi) +
{1 \over 2} (E_{R}\Gamma^{a}E_{R})(\Gamma_{a}D_{L}\phi)^{\hat
\alpha}
\label{2.26}
\end{eqnarray}
To conclude this section
let us recall some identities, which will be useful later
\begin{eqnarray}
   \Delta_{\hat \alpha}\Delta_{\beta}\phi = -
\Delta_{\beta}\Delta_{\hat \alpha}\phi = - {1 \over 2} (\Gamma^{c}M
\Gamma_{c})_{\beta \hat \alpha}
\label{2.23}
\end{eqnarray}
\begin{eqnarray}
\Delta_{[\alpha}\Delta_{ \beta]}\phi - \Delta_{\alpha}\phi \Delta_{\beta}\phi
 + {1 \over {4}}(\Gamma^{abc})_{\alpha \beta}H_{abc} = 0
\nonumber
\end{eqnarray}
\begin{eqnarray}
\Delta_{[\hat \alpha}\Delta_{ \hat \beta ]}\phi - \Delta_{\hat \alpha}
\phi \Delta_{\hat \beta}\phi -
 {1 \over {4}}(\Gamma^{abc})_{\hat \alpha \hat \beta}H_{abc} = 0
\label{2.24}
\end{eqnarray}
These identites follow by considering the sector (0,3) of the Bianchi identity
(\ref{2.10}) (that is the terms proportial to the product of three $E^{
\underline{\alpha}}$ ), using (\ref{2.17}) and computing $\Delta T^{
\underline{\alpha}}$ using (\ref{2.15}), (\ref{2.16}).
$\Delta_{\underline{\alpha}}$ are the spinor-like tangent space components of
the covariant diferential $\Delta$.
\section{BRS Algebra and Pure Spinors Constraints}

As already noted, the method of the Extended Differential Algebra proposed in
\cite{tor1} to derive BRST transformations
and pure spinor constraints is a generalization of a procedure \cite{Stora},
\cite{Bon1} well known for Yang-Mills theories, which
received a large
number of applications, as for instance in the  descent equations for the
consistent  anomalies in gauge theories \cite{Zum}, \cite{Bon2},  \cite{MSZ}
or  in some
treatments of topological twisted theories \cite{Ansel}, \cite{Baul}.

Let us describe the case of Yang-Mills theory to illustrate the
method. The Yang-Mills algebra involves the connection $A \in Lie G
$ valued in the Lie algebra of the gauge group G, with curvature $F
= dA + AA $ and Bianchi identity $\Delta F = dF + aF - FA = 0 $. The
extended algebra is obtained by replacing $ d $ with $\hat d = d +
\cal{S}$ and $A$ with $\hat A = A + c $ where $\cal{S}$ is the BRS
differential and $c \in Lie G $ is the ghost. The extended curvature
and the extended Bianchi indetities are defined by the same
algebraic relations as for the unextended ones. The extended
curvature is constrained to be equal to the initial one
\begin{eqnarray}
\hat F = \hat d \hat A + \hat A \hat A = F
\label{YM}
\end{eqnarray}
so that, expanding  in ghost number, eq.(\ref{YM}) is an identity 
in the sector with ghost number $ n_{gh}=0$. In the
sector with $ n_{gh} = 1 $ one has $ {\cal{S}} A = - d c - c A - A c
\equiv - \Delta c $, the BRS transformation of $A$ and in the sector
with $ n_{gh} = 2 $ one has $
{\cal{S}} c = - c c $, the BRS transformation of the ghost $c$.\\
Stora \cite{Stora} calls  eq. (\ref{YM}) the ``Russian formula'' (
see also
\cite{Bon1} for a ``superspace'' derivation of this formula).

This approach can be generalized to supergravity theories. The method
 consists of extending the (Free) Differential Algebra of supergravity by
replacing the differential $d$ with $d + {\cal S}$ where $ {\cal S}$
is the (full)  BRST differential
 and by adding to each gauge (super)-form an associated ghost. Since this
extension is purely algebraic the definitions of torsions and  curvatures and
 their  parametrizations remain the same for the extended objects.

 In \cite{tor1}  all  ghosts related to bosonic gauge
symmetries are set to zero (so that only the ghosts $\lambda^{\underline{
\alpha}}$, related to supersymmetry, are kept), resulting in a constrained
BRST algebra.

Our proposal differs from that of \cite{tor1} in two respects.

The first difference is that we will keep non--zero not only the
ghosts $\lambda^{\underline{\alpha}}$ related to supersymmetry but
also the ghosts $\sigma_{0}$, $\sigma_{1}$, $\sigma_{2}$ and
$\psi^{ab}
\equiv \psi^{[ab]}$ related to the gauge transformations of $C_{1}$,
$B_{2}$, $C_{3}$ and to Lorentz transformations respectively.

However we shall maintain that the ghosts $\lambda^{a}$, related to
translations, vanish. The reason for taking $\lambda^{a} = 0 $ is
that this condition is nothing else but the main  constraint of the
superembedding approach \cite{Super}, \cite{Y-form1}. The
superembedding approach provides an alternatrive description of the
Green-Schwarz approach where the k-symmetry of the Green-Schwarz
formulation is reinterpreted as worldsheet supersymmetry. Its main
feature is the requirment that the components of the pull-back of
the supervielbeins $E^{a}$ along the odd dimensions of the
superworldsheet, vanish. On the other hand, from the point of view
of  superembedding ,
 the ghosts $\lambda^{A}$ associated to the supervielbeins $E^{A}$ can be
identified with the pull-back of $E^{A}$ along a Grassmann--odd
dimension, parametrized by $\eta$, of the super worldsheet,
$\lambda^{A} =
\partial_{\eta}Z^{M}E_{M}{}^{A}$ \cite{Y-form1}, so that $
\lambda^{a} = \partial_{\eta}Z^{M}E_{M}{}^{a} = 0 $ is just the main
requirement of the superembedding approach.

The second difference is motivated by the fact that the BRST transformations
relevant for the pure spinor approach are those related to the ghosts
 $\lambda^{\underline{\alpha}}$. Therefore we split the full BRST
differential $\cal{S}$ as $$ {\cal{S}} = s + \delta $$ where
$ s = s_{L} + s_{R} $ is the BRST differential related to the ghosts
$\lambda^{\underline{\alpha}} \equiv (\lambda_{L},\lambda_{R})$ and $\delta$
generates the BRST  Lorentz  and gauge transformations with gauge parameters
given by the corresponding ghosts.

Now we can define the extended (hatted) quantities:
\begin{eqnarray}
 \hat E^{a} = E^{a}
\label{3.02}
\end{eqnarray}
$$ \hat E_{L}{}^{\alpha} = E_{L}{}^{\alpha} + \lambda_{L}{}^{\alpha} $$
\begin{eqnarray}
\hat E_{R}{}^{\hat \alpha} = E_{R}{}^{\hat \alpha} + \lambda_{R}{}^{
\hat \alpha}
\label{3.03}
\end{eqnarray}
As for the extension of  superforms other that $E^{A}$  (which can be
expressed on the basis of the supervielbeins $E^{A}$) there are two
possibilities: or one extends the supervielbeins themselves before adding the
ghosts or one keeps the supervielbeins unextended. Therefore one has
\begin{eqnarray}
 \hat \Omega_{ab} =\hat E^{C} \Omega_{C}{}^{ab} + \psi^{ab} \equiv
\Omega^{ab} + \tilde \psi^{ab}
\label{3.04}
\end{eqnarray}
\begin{eqnarray}
 \hat B_{2}= \hat E^{A}\hat E^{B}B_{BA} + \sigma_{1} \equiv  B_{2} +
\tilde \sigma_{1}
\label{3.05}
\end{eqnarray}
\begin{eqnarray}
 \hat C_{1} =\hat E^{A} C_{A} + \sigma_{0} \equiv   C_{1} + \tilde \sigma_{0}
\label{3.06}
\end{eqnarray}
 \begin{eqnarray}
 \hat C_{3} =  \hat E^{A}\hat E^{B}\hat E^{C}C_{CBA} + \sigma_{2} \equiv
C_{3} + \tilde \sigma_{2}
\label{3.1}
\end{eqnarray}
Of course there is a relation between  the ghosts and the ghosts tilded, that
follow from (\ref{3.04})-(\ref{3.1}), given (\ref{3.03}). Indeed
 $$\tilde \psi^{ab} = \psi^{ab} + i_{\lambda}\Omega^{ab}, $$
$$\tilde \sigma_{0} = \sigma_{0} + i_{\lambda}C_{1}, $$
\begin{eqnarray}
\tilde \sigma_{1} = \sigma_{1} + i_{\lambda}B_{2} +
 {1 \over 2} i_{\lambda}{}^{2}B_{2},
\label{3.07}
\end{eqnarray}
\begin{eqnarray}
\tilde \sigma_{2} = \sigma_{2} + i_{\lambda}C_{3} +
 {1 \over 2} i_{\lambda}{}^{2}C_{3} +  {1 \over 6} i_{\lambda}{}^{3}C_{3}.
\label{3.08}
\end{eqnarray}

$i_{\lambda}$ acting on a superform denotes the contraction of this superform
with $\lambda^{\underline{\alpha}}$.

One should notice that,
since $B_{2}$ and $C_{3}$ are superforms of degree higher than one,
$\sigma_{1}$ and $\sigma_{2}$ contain in principle ghosts of ghosts that is
$ \sigma_{1} = \sigma_{1}^{1} + \sigma_{0}^{2} $ and
$ \sigma_{2} = \sigma_{2}^{1} + \sigma_{1}^{2} + \sigma_{0}^{3}$
or  $\tilde \sigma_{1} =\tilde \sigma_{1}^{1} +\tilde \sigma_{0}^{2}
 $ and $\tilde \sigma_{2} = \tilde \sigma_{2}^{1} +\tilde \sigma_{1}^{2} +
\tilde \sigma_{0}^{3}$ where $\sigma_{i}^{p}$ and
$\tilde \sigma_{i}^{p}$ have ghost number p and form degree i. The
definition of $\tilde \sigma_{i}^{p}$ is obvious from (\ref{3.07}),
(\ref{3.08}) but notice in particular that $\tilde \sigma_{0}^{2}$
receives a  contribution also from $i_{\lambda}
\sigma_{1}$ and $\tilde \sigma_{1}^{2}$ and  $\tilde \sigma_{0}^{3}$ from
 $i_{\lambda}\sigma_{2}$ and  ${1 \over 2} i_{\lambda}^{2}\sigma_{2}$ .

Using these two options, we can define in the two ways the BRST
differential
\begin{eqnarray}
 \hat d = d+ s + \delta \equiv d + \tilde s + \tilde \delta.
\label{3.01}
\end{eqnarray}
In the first option realized by $\delta$, one assumes that $\delta$
induces infinitesimal Lorentz transformations with Lorentz parameter $\psi^{
ab}$ acting on Lorentz tensors, Lorentz connections and relative ghosts and
 gauge transformations with parameters $\sigma_{i}$, i =( 0,1,2), acting on
 $C_{1}$, $B_{2}$ and $C_{3}$ and relative ghosts. In particolar
 $$\delta C_{1} +  d \sigma_{0} = 0 = \delta B_{2} + d\sigma_{1}^{1} = 0, $$
$$ \delta  C_{3} +d \sigma_{2}^{1} + B_{2} d \sigma_{0} + d \sigma_{1}^{1}
C_{1} = 0 , $$ $$ \delta \Omega^{ab} + \Delta \psi^{ab}= 0.$$
Moreover
$$ \delta \sigma_{1}^{1} + d \sigma_{0}^{2}= 0 =
\delta \sigma_{2}^{1} + d \sigma_{1}^{2}, $$ $$ \delta \sigma_{1}^{2} + d
\sigma_{0}^{3} = 0 = \delta \sigma_{0}^{3} $$
 and
$$\delta \psi^{ab} = \psi^{a}{}_{c}\psi^{c b}.$$
Notice that $s + \delta$ is nilpotent and anticommutes with the differential
$d$. Moreover $\delta^{2}= 0$ and $d \delta + \delta d = 0 = s \delta +
\delta s $.
In this case  the BRST differential $s$ is nilpotent. However, just
to assure nilpotency, $s$, acting on non invariant quantities,
induces  Lorentz transformations with parameter  $
i_{\lambda}\Omega^{ab} $ and gauge transformations with parameters 
 $ i_{\lambda}C_{1}$, $i_{\lambda}B_{2}$
and $i_{\lambda}C_{3}$.

In the second option realized by $\tilde \delta$, rewriting  $$s +
\delta =
 \tilde s + \tilde \delta $$ as anticipated in (\ref{3.01}), one
assumes that $\tilde \delta$ acts as $ \delta $ but with the Lorentz and gauge
 parameters $\psi^{ab}$ and $\sigma_{i}$  replaced by   $\tilde \psi^{ab}$
and $\tilde \sigma_{i}$. In this case the non covariant Lorentz and gauge
transformations induced by $s$ are absorbed by $\tilde \delta$ and $\tilde s$
is the  covariant BRST differential. Now $\tilde \delta$ does not
anticommute with $\tilde s$ and therefore $\tilde s $ is no longer nilpotent.
For instance, acting on the Lorentz vector  $V^{a}$
\begin{eqnarray}
 \tilde s^{2} V^{a} = - {1 \over 2} V^{b}i_{\lambda}{}^{2}R_{b}{}^{a}
\label{3.09}
\end{eqnarray}
However, acting on Lorentz scalar and gauge invariant superfields,
$\tilde s$ coincides with $ s $ and is nilpotent.

Since s is always nilpotent, the nilpotent BRST charge Q, relevant
to the pure spinor approach, generates the transformations induced
by $ s $ (not $\tilde s$).

Now one can  define the torsions and curvatures of the hatted
quantities as done in  (\ref{2.9})-(\ref{2.14}) for the unhatted
ones. Since the hatted superforms have the same algebraic properties
as the unhatted ones, they obey the same  Bianchi identities and can
be subject to the same rheonomic parametrization, namely

\begin{eqnarray}
\hat T^{a} =\hat \Delta \hat E^{a} = {1 \over 2}(\hat E_{L}\Gamma^{a}
\hat E_{L}) + {1 \over 2} (\hat E_{R}\Gamma^{a}\hat E_{R})
\label{3.2}
\end{eqnarray}
\begin{eqnarray}
\hat H_{3}= \hat d \hat B_{2} = - \hat E^{a}[(\hat E_{L}\Gamma_{a}\hat E_{L})
- (\hat E_{R}\Gamma_{a}\hat E_{R}) ]+ \hat E^{a}\hat E^{b} \hat E^{c}H_{abc}
\label{3.3}
\end{eqnarray}
\begin{eqnarray}
\hat T_{L}^{\alpha}=\hat \Delta \hat E_{L}^{\alpha} = {1 \over 4}[
\hat E_{L}^{\alpha}
(\hat E_{L} D_{R}\phi) + {1 \over 2}(\Gamma^{ab}\hat E_{L})^{
\alpha}( D_{R}\phi\Gamma_{ab}\hat E_{L})] - {3 \over 4} \hat E^{c}(\Gamma^{ab}
\hat E_{L})^{\alpha} H_{c a b} \nonumber
\end{eqnarray}
\begin{eqnarray}
  + \hat E^{c}(M \Gamma_{c}\hat E_{R})^{\alpha}+
\hat E^{b}\hat E^{c}T_{L{} b c}{}^{\alpha}
\label{3.4}
\end{eqnarray}
\begin{eqnarray}
 \hat T_{R}^{\hat\alpha}=\hat \Delta \hat E_{R}^{\hat \alpha} = {1 \over 4}[
\hat E_{R}^{\hat\alpha}(\hat E_{R}D_{L}\phi) + {1 \over 2}
 (\Gamma^{ab}\hat E_{R})^{\hat \alpha}(D_{L}\phi\Gamma_{ab}\hat E_{R})] +
 {3 \over 4} \hat E^{c}(\Gamma^{ab}\hat E_{R})^{\hat \alpha}H_{cab} \nonumber
\end{eqnarray}
\begin{eqnarray}
 - \hat E^{c}
(\hat E_{L}\Gamma_{c}M )^{\hat \alpha}+ \hat E^{b}\hat E^{c}T_
{R{} b c}{}^{\hat \alpha}
\label{3.5}
\end{eqnarray}
$$ \hat R^{ab}=  \hat d \hat \Omega^{ab} - \hat \Omega^{a}{}_{c}
\hat \Omega^{c b}
 = {3 \over 2}[(\hat E_{L}\Gamma^{c}\hat E_{L}) - (\hat E_{R}\Gamma^{c}\hat E_{
R})]H^{cab} + 2 (\hat E_{L}\Gamma^{[a}M\Gamma^{b]}\hat E_{R})$$
\begin{eqnarray}
+ \hat E^{c}(
(\hat E_{L}\Theta_{L{} c}{}^{a b}) +
(\hat E_{R}\Theta_{R{} c}{}^{a b})) +\hat E^{c}\hat E^{d}R_{cd}{}^{ab}
\label{3.6}
\end{eqnarray}
\begin{eqnarray}
\hat G_{2}= \hat d \hat C_{1} = - e^{\phi}(\hat E_{R}\hat E_{L}) -
e^{\phi}\hat E^{a}[(\hat E_{L} \Gamma_{a}D_{L}\phi) -
 (\hat E_{R}\Gamma_{a}D_{R}\phi)] + \hat E^{a}\hat E^{b}G_{ab}
\label{3.7}
\end{eqnarray}
\begin{eqnarray}
\hat G_{4}= \hat d \hat C_{3} +\hat B_{2}\hat d \hat C_{1}  = e^{\phi}
\hat E^{a}\hat E^{b}(\hat E_{L}\Gamma_{ab}\hat E_{R})  \nonumber
\end{eqnarray}
\begin{eqnarray}
+{1 \over 3} e^{\phi}\hat E^{a}\hat E^{b}
\hat E^{c}(( D_{L}\phi\Gamma_{abc}\hat E_{L}) + (D_{R}\phi \Gamma_{abc}
\hat E_{R})) + \hat E^{a}\hat E^{b}\hat E^{c}\hat E^{d}G_{abcd}
\label{3.8}
\end{eqnarray}
In the sector with zero ghost number the equations  (\ref{3.2})-(\ref{3.8})
reproduce the equations  (\ref{2.9})-(\ref{2.14}).

In the sector with ghost number one, following the second option,
and noticing that in the left hand sides of eqs.
(\ref{3.2})-(\ref{3.8}) the action of $\tilde \delta $  cancels the
tilded ghosts, these equations  give
\begin{eqnarray}
\tilde s E^{a} = (\lambda_{L}\Gamma^{a}E_{L}) +  (\lambda_{R}\Gamma^{a}E_{R})
\label{3.10}
\end{eqnarray}
\begin{eqnarray}
\tilde s B_{2} = -2 E^{a}[(\lambda_{L}\Gamma_{a}E_{L})-(\lambda_{R}
\Gamma_{a}E_{R})]
\label{3.11}
\end{eqnarray}

$$
\tilde s E_{L}{}^{\alpha} = -\Delta \lambda_{L}{}^{\alpha} +{1 \over 4} [
\lambda_{L}{}^{\alpha}(E_{L}D_{R}\phi) + {1 \over 2} (\Gamma^{ab}
\lambda_{L})^{\alpha}(D_{R}\phi\Gamma_{ab}E_{L})]$$
\begin{eqnarray}
 +{1 \over 4} [E_{L}{}^{\alpha}(
\lambda_{L}D_{R}\phi) +{1 \over 2}(\Gamma^{ab}
E_{L})^{\alpha}(D_{R}\phi\Gamma_{ab}\lambda_{L})]
 - {3 \over 4} E^{c}H_{cab}(\Gamma^{ab})
^{\alpha}{}_{\beta} \lambda_{L}{}^{\beta} + (M E^{c}\Gamma_{c}\lambda_{R})
^{\alpha}
\label{3.12}
\end{eqnarray}
$$
\tilde s E_{R}{}^{\hat \alpha} = -\Delta \lambda_{R}{}^{\hat \alpha} +
{1 \over 4} [\lambda_{R}{}^{\hat\alpha}(E_{R}D_{R}\phi) + {1 \over 2}
(\Gamma^{ab}\lambda_{R})^{
\hat\alpha}(D_{R}\phi\Gamma_{ab}E_{R})] $$
\begin{eqnarray}
 +{1 \over 4} [E_{R}{}^{\hat\alpha}(\lambda_{R}D_{R}\phi) +
{1 \over 2} (\Gamma^{ab}E_{R})^{\hat\alpha}(D_{R}\phi\Gamma_{ab}\lambda_{R})] +
{3 \over 4} E^{c}H_{cab}(\Gamma^{ab})
^{\hat\alpha}{}_{\hat \beta} \lambda_{R}{}^{\hat \beta} - (\lambda_{L} E^{c}
\Gamma_{c}M)^{\hat \alpha}
\label{3.13}
\end{eqnarray}
$$\tilde s\Omega^{ab}  = 3 [(E_{L}\Gamma^{c}\lambda_{L}) - (E_{R}\Gamma_{c}
\lambda_{R})]H^{cab} + 2(\lambda_{L}\Gamma^{[a} M \Gamma^{b]}E_{R}) + 2
(E_{L}\Gamma^{[a} M \Gamma^{b]}\lambda_{R})$$
\begin{eqnarray}
 - E^{c}[(\lambda_{L}\Theta_{L{} c}{}^{a b}) +
(\lambda_{R}\Theta_{R{} c }{}^{a b})]
\label{3.14}
\end{eqnarray}
\begin{eqnarray}
\tilde s C_{1} = - e^{\phi}[ (\lambda_{L}E_{R}) + (\lambda_{R} E_{L})] - e^{-\phi}
E^{a}[ (\lambda_{L}\Gamma_{a} D_{L}\phi) - (\lambda_{R}\Gamma_{a}D_{R}\phi)]
\label{3.15}
\end{eqnarray}
$$
\tilde s C_{3} = - B_{2}\tilde s C_{1} + e^{\phi}E^{a}E^{b}[(\lambda_{L}\Gamma_{ab}
E_{R}) +
(E_{L}\Gamma_{ab}\lambda_{R})]$$
\begin{eqnarray}
 - {1 \over 3} E^{a}E^{b}E^{c}e^{\phi}[ (\lambda_{L}
\Gamma_{abc}D_{L}\phi) + (\lambda_{R}\Gamma_{abc}D_{R}\phi)]
\label{3.16}
\end{eqnarray}
Equations  (\ref{3.10})-(\ref{3.16}) are the BRST transformations of super
vielbeins, Lorentz connection and gauge forms, derived as in \cite{tor1}.

The sector with ghost number two is the most interesting for our purposes
since it gives the BRST trasformations of the ghosts and the ghost constraints.

Before discussing this sector it is convenient to do the mild assumption
 that
\begin{eqnarray}
\tilde s \tilde \sigma_{1}^{1} = 0
\label{3.001}
\end{eqnarray}
Later on we will see that this assumption is quite natural and
almost implied by the formalism.

Since $\lambda^{a} = 0 $, from eq. (\ref{3.2}) one has
\begin{eqnarray}
 0 =  (\lambda_{L}\Gamma^{a}\lambda_{L}) + (\lambda_{R}\Gamma^{a}
\lambda_{R})
\label{3.17}
\end{eqnarray}
With the assumption (\ref{3.001}), eq. (\ref{3.3}) gives
\begin{eqnarray}
0 = - E^{a}[(\lambda_{L}\Gamma^{a}\lambda_{L})- (\lambda_{R}\Gamma^{a}
\lambda_{R})]
\label{3.18}
\end{eqnarray}
Since the vector supervielbeins are generic, equations (\ref{3.17}),
(\ref{3.18}) imply the pure spinor constraints
\begin{eqnarray}
(\lambda_{L}\Gamma^{a}\lambda_{L}) = 0 =  (\lambda_{R}\Gamma^{a}\lambda_{R})
\label{3.9}
\end{eqnarray}
Using the extended version of the definition (\ref{2.25}),
at the level of $n_{gh} = 2 $,
 eqs. (\ref{3.4}) and (\ref{3.5}) imply
\begin{eqnarray}
\tilde s \lambda_{L}{}^{\alpha}=  \lambda_{L}{}^{\beta}\lambda_{L}{}^{\gamma}
X_{\beta\gamma}{}^{\alpha}
\nonumber
\end{eqnarray}
\begin{eqnarray}
\tilde s \lambda_{R}{}^{\hat \alpha}=  \lambda_{R}{}^{\hat \beta}
\lambda_{R}{}^{\hat \gamma}X_{\hat \beta \hat \gamma}{}^{\hat \alpha}
\label{3.190}
\end{eqnarray}
or, taking into account eqs.(\ref{2.26}) and (\ref{3.9}),
\begin{eqnarray}
\tilde s\lambda_{L}{}^{\alpha}= - \lambda_{L}{}^{\alpha}(\lambda_{L}D_{R}\phi)
\nonumber
\end{eqnarray}
\begin{eqnarray}
\tilde s\lambda_{R}{}^{\hat \alpha}= - \lambda_{R}{}^{\hat \alpha}
(\lambda_{R}D_{L}\phi)
\label{3.19}
\end{eqnarray}
Finally, from (\ref{3.6}),(\ref{3.7}) and (\ref{3.8}) one has
\begin{eqnarray}
 \tilde s \tilde \psi^{ab} =
(\lambda_{L}\Gamma^{[a} M \Gamma^{b]}\lambda_{R}) + {3 \over 2} [(
\lambda_{L}\Gamma_{c}\lambda_{L})- (\lambda_{R}\Gamma_{c}\lambda_{R})]H^{cab}
\label{3.20}
\end{eqnarray}
\begin{eqnarray}
\tilde s \tilde \sigma_{0} = - e^{\phi}(\lambda_{R}\lambda_{L})
\label{3.21}
\end{eqnarray}
\begin{eqnarray}
\tilde s \tilde \sigma_{2}^{1} +  \tilde \sigma_{1}^{1}
\tilde s C_{1} + \tilde \sigma_{0}^{2}G_{2} = e^{\phi} [ E^{a}E^{b}(
\lambda_{L}\Gamma_{ab}\lambda_{R}) + B_{2}(\lambda_{L}\lambda_{R})]
\label{3.22}
\end{eqnarray}
In \cite{tor1}
the constraints for the $\lambda$'s are given by (\ref{3.17}), (\ref{3.18})
with $E^{a}$ restricted to the world sheet of the string together with
 two further
constraints given by the r.h.s. of  equations (\ref{3.21}), (\ref{3.22}).
  In our approach, due to  the
presence of the ghosts $ \tilde \sigma_{0}$ , $\tilde \sigma_{2}$ and $
\tilde \psi^{ab}$,
 these last constraints are avoided as well as the constraint given by the
r.h.s. of (\ref{3.20}) in the absence of the ghost $\tilde \psi^{ab}$.

Now let us show that the assumption (\ref{3.001}) is (almost) implied by the
extended Free Differential Algebra under  consideration.

Without the assumption (\ref{3.001}), equation (\ref{3.18}) becomes
\begin{eqnarray}
\tilde s \tilde \sigma_{1}^{1} = -E^{a}[
(\lambda_{L}\Gamma^{a}\lambda_{L})- (\lambda_{R}\Gamma^{a}\lambda_{R})]
\label{3.002}
\end{eqnarray}
Now consider the extended Bianchi identity
$$ \hat d \hat G_{4} =  \hat H_{3} \hat G_{2} $$ which is indeed
automatically
satisfied given (\ref{3.3}), (\ref{3.7}), (\ref{3.8}). In the sector of ghost
number four it gives
\begin{eqnarray}
e^{\phi} E^{a}[(\lambda_{L}\Gamma^{b}\lambda_{L})+ (\lambda_{R}\Gamma^{b}
\lambda_{R})](\lambda_{L}\Gamma^{ab}\lambda_{R}) = e^{\phi}E^{a}[
(\lambda_{L}\Gamma^{a}\lambda_{L})- (\lambda_{R}\Gamma^{a}\lambda_{R})](
\lambda_{L}\lambda_{R})
\label{3.003}
\end{eqnarray}
Then, from (\ref{3.17}) and assuming
\begin{eqnarray}
(\lambda_{L}\lambda_{R}) \ne 0,
\label{3.004}
\end{eqnarray}
  equations (\ref{3.18}) and (\ref{3.001}) follow.
The other possible solution of (\ref{3.003}),
$(\lambda_{L}\lambda_{R}) = 0 $, is the one relevant to the D=11,
supermembrane \cite{Ber7}, \cite{Fre2}. Equation (\ref{3.003}) is
quite obvious since it is nothing else but the Fierz identity in
eleven dimensions $$ (\Gamma^{\underline{a} \underline{b}})_ {(
\underline{\alpha} \underline{\beta}}(\Gamma_{\underline{b}})_
{\underline{\gamma} \underline{\delta})} = 0 , $$ reduced to 10
dimensions and saturated with $\lambda^{\underline{\alpha}}$,
$\lambda^{\underline{\beta}}$, $\lambda^{\underline{\gamma}}$,
$\lambda^{\underline{\delta}}$

Given the equations (\ref{3.10})-(\ref{3.22}) one should verify the
nilpotency of the BRST charge $Q$ acting on these superforms and
ghosts. This can be done systematically starting from the relevant
Bianchi identities but in the most of the cases it is very easy to
perform the checks directly.

We conclude this section by deriving a useful property of the left
handed and right handed BRS differential of the bispinor $ P^{\alpha
\hat \beta}$. Using (\ref{3.18}), equation (\ref{3.20}) can be
written as $ (\tilde s +
\tilde \delta)\tilde \psi^{ab} = \tilde \psi^{a}{}_{c}\tilde \psi^{cb} +
(\lambda_{L}\Gamma^{[a}M\Gamma^{b]}\lambda_{R})$. Then the
nilpotency of $(s+\delta) = (\tilde s + \tilde \delta ) $ yields the
identity $$ \tilde s (\lambda_{ L}\Gamma^{[a} e^{\phi}P
\Gamma^{b]}\lambda_{R}) = 0 $$ where $ e^{\phi} P^{\alpha \hat
\beta}$ is defined in (\ref{2.22}). Taking into account
(\ref{3.19}), this equation  gives
\begin{eqnarray}
(\lambda_{L}\Gamma^{[a}[(\lambda_{L}{}^{\alpha}\Delta_{ \alpha} P +
\lambda_{R}{}^{\hat \alpha}\Delta_{\hat \alpha} P)]
\Gamma^{b]}\lambda_{R}) = 0
\label{3.23}
\end{eqnarray}
Equation (\ref{3.23}) also follows from the Bianchi identity
 $$ \hat \Delta \hat R^{ab} = 0$$
   in the sector
with ghost number three.

If one defines
\begin{eqnarray}
\tilde s_{L} P^{\beta \hat \gamma}\equiv \lambda_{L}{}^{\alpha}\Delta_{\alpha}
P^{\beta \hat \gamma} =
\lambda_{L}{}^{\alpha}C_{L{} \alpha}{}^{\beta \hat \gamma}
\nonumber
\end{eqnarray}
\begin{eqnarray}
\tilde s_{R} P^{\beta \hat \gamma}\equiv \lambda_{R}{}^{\hat \alpha}\Delta_{
\hat \alpha} P^{\beta \hat \gamma} =
\lambda_{R}{}^{\hat \alpha}C_{R{} \hat \alpha}{}^{\beta \hat \gamma}
\label{3.24}
\end{eqnarray}
equation (\ref{3.23}) implies
\begin{eqnarray}
 C_{L{} \alpha}{}^{\beta \hat \gamma} = \delta_{\alpha}{}^{\beta}C_{L}{}^{
\hat \gamma} + {1 \over 4} (\Gamma^{ab})_{\alpha}{}^{\beta}C_{L{} ab}{}^{
\hat \gamma}
\nonumber
\end{eqnarray}
\begin{eqnarray}
C_{R{} \hat\alpha}{}^{\beta \hat\gamma} = \delta_{\hat \alpha}{}^{
\hat \gamma}C_{R}{}^{\beta} +{1 \over 4}(\Gamma^{ab})_{\hat \alpha}{}^{
\hat \gamma}C_{R{} ab}{}^{ \beta}
\label {3.25}
\end{eqnarray}

Now note that a generic matrix $ Y_{\alpha}{}^{\beta} $ can be
decomposed as follows:
$$ Y_{\alpha}{}^{\beta} = Y^{(0)}\delta_{\alpha}{}^{\beta} + Y^{(2)}_{ab}
(\Gamma^{ab})_{\alpha}{}^{\beta} +  Y^{(4)}_{abcd}
(\Gamma^{abcd})_{\alpha}{}^{\beta}. $$ If the term $Y^{(4)}_ {abcd}
$ is absent we shall say that $Y_{\alpha}{}^{\beta}$ is {\it Lorentz
and Weyl valued } or in short {\it LW-valued} . The same is valid
for the matrix  $Y_{\hat \alpha}{}^{\hat \beta}$. For instance,
according to (\ref{3.24}), $C_{L \alpha}{}^{\beta \hat \gamma}$ and
$ C_{R
\hat
\alpha} {}^{\beta \hat\gamma}$ are LW-valued with respect to the
indices $\alpha,\beta$ and $\hat \alpha,
\hat \gamma$ respectively.

This result will be important in the next section to assure the consistency
of the pure spinor action.

\section{ Pure Spinor Action}
In this section we derive the pure spinor action, in two different ways
that will be described in the subsections 4.1 and 4.2 respectively.
As a result we recover  the action first obtained in \cite{BH}.

Before doing that, some preliminary considerations are in order. The
ghosts $\sigma$'s and $\psi $ do not appear in the  action, while
one must add to the superspace coordinates $Z^{M}$ and the
 ghosts $\lambda^{\underline{\alpha}}$ the antighosts $$\omega_{\underline{\alpha}} =
 (\omega_{R \alpha},\omega_{ L \hat \alpha}) $$ with ghost number
$n_{g}= -1 $,  which are the
 conjugate momenta of $\lambda^{\underline{\alpha}}$ , and the fields
$$d_{\underline{\alpha}} = ( d_{R \alpha}, d_{L \hat \alpha})$$
that involve the conjugate momenta of $Z^M$ and are essentially the BRS
partners of $\omega_{\underline{\alpha}}$. From the worldsheet point of view,
$\omega_{R}$, $d_{R}$ and $\omega_{L}$, $d_{L}$ are right-handed and
left-handed chiral  fields respectively.

Since, as a consequence of the pure spinor constraints, $\lambda^{
\underline{\alpha}}$ contains 11  + 11 degree of freedom, also $\omega_{
\underline{\alpha}}$ should contain 11 + 11 independent components. This is
realized by assuming that the pure spinor action is invariant under the
$\omega$-gauge symmetry
\begin{eqnarray}
 \delta^{(\omega)} \omega_{R/L} = \Lambda_{R/L}^{a}(\Gamma_{a}\lambda_{L/R})
\label{4.10}
\end{eqnarray}
 where $\Lambda_{R/L}^{a}$ are gauge parameters.

The $d_{\underline{\alpha}}$ appear
in the nilpotent BRS charge \footnote{Here and in the following $\oint$
denotes $\int
d\sigma_{+}$ or $\int d\sigma_{-}$ according to the case where $\sigma_{\pm}$
are worldsheet light-cone coordinates.}
$$ Q =  Q_{L} + Q_{R} = \oint \lambda^{\underline{\alpha}}d_{
\underline{\alpha}}$$
that is $$ Q_{L} =\oint (\lambda_{L}d_{ R}),$$ $$Q_{R} =  \oint
(\lambda_{R} d_{L}).$$ under  which  the action must be invariant.
In order to specify $ Q$ and prove its nilpotency one needs the
expression for $d_{\underline{\alpha}}$. As has already been noted,
$Q$ generates the transformations induced by the BRS differential $
s $ and a suggestion to obtain an ansatz for  $d_{
\underline{\alpha}}$ comes from the action of $ s$ on superfields and ghosts.
In particular from (\ref{3.190}) one has
\begin{eqnarray}
s \lambda^{\underline{\alpha}} = \lambda^{\underline{\beta}}\lambda^{
\underline{\gamma}}
 [\Omega_{\underline{\beta}\underline{\gamma}}{}^{\underline{\alpha}} +
 X_{\underline{\beta}\underline{\gamma}}{}^{\underline{\alpha}}]
\label{4.0}
\end{eqnarray}
Since the  BRS trasformations
of $\lambda^{\underline{\alpha}}$ do not vanish,
 $d_{\underline{\alpha}}$ must contain  terms linear in $\omega$  in order to
reproduce (\ref{4.0}). Therefore a general form of $ d_{\underline{\alpha}}$
is expected to be
$$ d_{R \alpha} = d^{(0)}_{R \alpha} + (\Omega_{\alpha \beta}{}^{\gamma} +
X_{\alpha \beta}{}^{\gamma}) \omega_{R \gamma}\lambda_{L}{}^{\beta} +
\Omega_{\alpha \hat \beta}{}^{\hat \gamma} \omega_{L \hat \gamma}\lambda_{R}{}
^{\hat \beta} $$
$$ d_{L \hat \alpha} = d^{(0)}_{L \hat \alpha} + (\Omega_{\hat \alpha \hat
\beta}
{}^{\hat \gamma} + X_{\hat \alpha \hat \beta}{}^{\hat \gamma}) \omega_{L \hat
\gamma}\lambda_{R}{}^{\hat \beta} + \Omega_{\hat \alpha \beta}{}^{ \gamma}
\omega_{R \gamma}\lambda_{L}{}^{\beta} $$
where $d^{(0)} $ does not depend on $\omega$ and $\lambda$ and $\Omega_{
\underline{\alpha} \underline{\beta}}{}^{\underline{\gamma}}$ are the
tangent space spinorial components of the Lorentz connection.There
is an ambiguity in the form of $
X_{\underline{\beta}\underline{\gamma}}{}^{\underline{\alpha}}$
since  (\ref{4.0}) specify only the component of this superfield
which is symmetric in $\underline{\beta},\underline{\gamma}$ and
modulo the pure spinor constraints. A convenient choice of $
X_{\underline{\alpha} \underline{\beta}}{}^{\underline{\gamma}}
\equiv (  X_{\alpha \beta}{}^{\gamma},  X_{\hat \alpha \hat
\beta}{}^{\hat \gamma} )$ that reproduces (\ref{4.0}) (i.e
(\ref{3.19})) and is LW-valued in $\underline{\beta},
\underline{\gamma}$ is
$$ X_{\alpha \beta}{}^{\gamma} = {1 \over 4} [ \delta_{\beta}{}^{\gamma}D_{R
\alpha} \phi + {1 \over 2} (\Gamma^{ab})_{\beta}{}^{\gamma}(\Gamma_{ab}D_{R}
\phi)_{\alpha} ]$$
\begin{eqnarray}
X_{\hat \alpha \hat \beta}{}^{\hat \gamma} = {1 \over 4} [ \delta_{\hat \beta}
{}^{\hat \gamma}D_{R \hat \alpha} \phi + {1 \over 2} (\Gamma^{ab})_{\hat
\beta}{}^{\hat \gamma}(\Gamma_{ab})_{\hat \alpha}{}^{\hat \delta}D_{R
\hat \delta}\phi ]
\label{4.1}
\end{eqnarray}
$d^{(0)}_{\underline{\alpha}}$ acting on superfields induces the tangent space
 derivatives $ D_{\underline{\alpha}} $
and
\begin{eqnarray}
 \lbrace   Q, (d^{(0)}_{R/L\underline{\alpha}}\rbrace
=  - (E_{\mp}^{a}\Gamma_{a}\lambda_{L/R})_{\underline{\alpha}} +
 2 \lambda^{\underline{\beta}}( 
\Omega_{ (\underline{\beta} \underline{\alpha})}{}^{\underline{\gamma}} +
X_{ (\underline{\beta} \underline{\alpha})}{}^{\underline{\gamma}})d_{
\underline{\gamma}}
\label{4.2}
\end{eqnarray}
where $E_{\pm}^{a}$ are the pullbacks of the vector-like vielbeins
$E^{a}$ on the worldsheet.

With this expression for $ d_{\underline{\alpha}}$ the BRS charge is
$$ Q = \oint \lambda^{\underline{\alpha}}[ d^{(0)}_{\underline{\alpha}} +
(\Omega_{\underline{\alpha} \underline{\beta}}{}^{\underline{\gamma}} +
X_{\underline{\alpha} \underline{\beta}}{}^{\underline{\gamma}})
\omega_{\underline{\gamma}}\lambda^{\underline{\beta}}] $$
and   $$ Q^2 = -  \oint \lambda^{\underline{
\alpha}}\lambda^{\underline{\beta}}\lambda^{\underline{\gamma}}
\tilde R_{\underline{\alpha}\underline{\beta} \underline{\gamma}}{}^{
\underline{\delta}}\omega_{\underline{\delta}} $$ where
$ \tilde R_{L \alpha}{}^{\beta}$ and $ \tilde R_{R \hat \alpha}{}^{\hat \beta}$
are the curvatures of the Lorentz connections $ \tilde \Omega_{L \alpha}{}^{
\beta} = \Omega_{\alpha}{}^{\beta} +  E_{L}^{\gamma} X_{\gamma \alpha}{}^{
\beta}$
and $ \tilde \Omega_{R \hat \alpha}{}^{\hat \beta} = \Omega_{\hat \alpha}{}^{
\hat \beta} + E_{R}^{\hat \gamma} X_{\hat \gamma \hat \alpha}{}^{\hat \beta}$

By an explicit computation of the l.h.s. of this equation and using
the identities (\ref{2.23}) and (\ref{2.24}) one can verify that $Q$
is indeed  nilpotent. {}From the expression for $ Q $ one can also
compute the BRS transformations of $\omega_{\underline{\alpha}}$ and
$ d_{\underline{\alpha}}$
\begin{eqnarray}
 s \omega_{\underline{\alpha}} = - d_{\underline{\alpha}} -
 \lambda^{\underline{\beta}}(\Omega_{\underline{\beta} \underline{\alpha}}
{}^{\underline{\gamma}} + X_{\underline{\beta} \underline{\alpha}}
{}^{\underline{\gamma}})\omega_{\underline{\gamma}}
\label{4.4}
\end{eqnarray}
\begin{eqnarray}
 s d_{ \underline{\alpha}} =
 - (E_{\mp}^{a}\Gamma_{a})_{\underline{\alpha} \underline{\beta}}\lambda^{
\underline{\beta}} + 2 \lambda^{\underline{\beta}} \tilde R_{
\underline{\alpha} \underline{\beta} \underline{\gamma}}{}^{
\underline{\delta}}
\lambda^{\underline{\gamma}}\omega_{\underline{\delta}} +  \lambda^{
\underline{\beta}}(\Omega_{\underline{\beta} \underline{\alpha}}
{}^{\underline{\gamma}} + X_{\underline{\beta} \underline{\alpha}}
{}^{\underline{\gamma}})d_{\underline{\gamma}}
\label{4.5}
\end{eqnarray}

The covariant form of equation (\ref{4.4}) is
\begin{eqnarray}
 \tilde s \omega_{\underline{\alpha}} = - d_{\underline{\alpha}} -
\lambda^{\underline{\beta}}X_{\underline{\beta} \underline{\alpha}}{}^{
\underline{\gamma}}\omega_{\underline{\gamma}},
\label{4.55}
\end{eqnarray}
that is
$$  \tilde s_{L/R} \omega_{R/L \underline{\alpha}} = - d_{R/L
\underline{\alpha}} -
\lambda_{L/R}^{\underline{\beta}}X_{\underline{\beta} \underline{\alpha}}{}^{
\underline{\gamma}}\omega_{R/L \underline{\gamma}}$$
$$ \tilde s_{R/L} \omega_{R/L \underline{\alpha}} = 0 $$
Eq. (\ref{4.55}) contains a term proportional to $
(\omega_{R/L}\Gamma_{abcd}\lambda_{L/R})$ which is non invariant
under (\ref{4.10}). It is useful to remark that in the covariant BRS
transformation  of $e^{\phi}\omega_{ R/L}$ this ill-behaved term is
removed. Indeed, using the identity
$$ \omega_{\underline{\alpha}}\lambda^{\underline{\beta}} = {1 \over {16}}
 [ \delta_{
\underline{\alpha}}{}^{\underline{\beta}} (\omega\lambda) - {1 \over 2}
(\Gamma^{ab})_{\underline{\alpha}}{}^{\underline{\beta}} (\omega \Gamma_{ab}
\lambda) + {1 \over {24}} (\Gamma^{abcd})_{\underline{\alpha}}{}^{\underline{
\beta}} (\omega \Gamma_ {abcd}\lambda)] , $$ one has $$ \lambda_{L/R}^{
\underline{\beta}}X_{\underline{\beta} \underline{\alpha}}{}^{
\underline{\gamma}}\omega_{R/L \underline{\gamma}} + (\lambda_{L/R}\Delta_{R/L}
\phi) \omega_{R/L \underline{\alpha}}  \equiv  \lambda_{L/R}^{
\underline{\beta}}Y_{\underline{\beta} \underline{\alpha}}{}^{
\underline{\gamma}}\omega_{R/L \underline{\gamma}} $$ where
$$ Y_{\beta \alpha}{}^{\gamma} = {1 \over 4} [ 3 \delta_{
\beta}{}^{\gamma} D_{\alpha}\phi - {1 \over 2} (\Gamma_{ab})_{
\beta}{}^{\gamma}(\Gamma_{ab}D_{R}\phi)_{\alpha} ] =  Y_{\alpha \beta}{}^{
\gamma} $$
\begin{eqnarray}
 Y_{\hat \beta \hat \alpha}{}^{\hat \gamma}  = {1 \over 4}
 [ 3 \delta_{\hat \beta}{}^{\hat \gamma}  D_{\hat \alpha}\phi - {1 \over 2}
 (\Gamma_{ab})_{\hat \beta}{}^{\hat \gamma}(\Gamma_{ab} D_{L}\phi)_{
\hat \alpha} ] =  Y_{\hat \alpha \hat \beta}{}^{\hat \gamma}
\label{4.12}
\end{eqnarray}
are LW-valued superfields with respect to $\beta,\gamma$ and $\hat
\beta,
\hat \gamma $ respectively and
\begin{eqnarray}
\tilde s_{L/R}(e^{\phi} \omega_{R/L \underline{\alpha}}) = - d_{ R/L
\underline{\alpha}} + \lambda_{L/R}^{ \underline{\beta}}\omega_{R/L
\underline{\gamma}} Y_{\underline{\beta}\underline{\alpha}}{}^{\underline{
\gamma}}
\label{4.11}
\end{eqnarray}
It is instructive to compute the action of $Q^2$ on
$\lambda^{\underline{\alpha}}$, $d_{\underline{\alpha}}$ and $\omega_{
\underline{\alpha}}$.

For $\lambda$ one  can confirm that $ Q $ applied to
$\lambda^{\underline{\alpha}}$ is nilpotent since $ s^{2} \lambda^{
\underline{\alpha}} = \lambda^{\underline{
\delta}}\lambda^{\underline{\beta}}\lambda^{\underline{\gamma}}
\tilde R_{\underline{\delta}\underline{\beta} \underline{\gamma}}{}^{
\underline{\alpha}} = 0 $.

At first sight, this is not true for $d_{\underline{\alpha}}$ since
taking the BRST differential of (\ref{4.5}), after some algebra, one obtains
\begin{eqnarray}
s^{2} d_{R/L\underline{\alpha}} =   -
(\Gamma^{a}\lambda_{L/R})_{\underline{\alpha}}
(\lambda_{R/L}\Gamma_{a})_{\underline{\beta}}[E_{\mp R/L}^{\underline{\beta}}
 \pm (s_{L/R}
( M_{R/L} \omega_{R/L})^{\underline{\beta}}]
\label{4.7}
\end{eqnarray}
where   $ M_{L} = M$,  $M_{R}$
 is the transpose of  $ M $ and M is defined in (\ref{2.22}). However, if we
assume that $E_{\mp R/L}$ satisfy
the field equations
\begin{eqnarray}
(E_{\pm R/L})^{\underline{\alpha}} \pm s_{L/R} (M_{R/L}^{
\underline{\alpha}\underline{\beta}} \omega_{R/L \underline{\beta}}) = 0,
\label{4.8}
\end{eqnarray}
$  s^{2}d_{R/L} $ vanishes on shell.
Afterward we shall prove that these field equations are  indeed satisfied
as variations of the pure spinor  action with respect to
$d_{\underline{\alpha}}$.

However $s^{2} \omega $ does not vanish  since from (\ref{4.4}) and
(\ref{4.5}) one has
\begin{eqnarray}
s^{2} \omega_{R/L} = ( E_{\mp}{}^{a}\Gamma_{a}\lambda_{L/R})
\label{4.9}
\end{eqnarray}
  This failure of nilpotency is a consequence of the $\omega$-gauge
transformation (\ref{4.10}). Indeed $s^2$,
acting on $\omega$, vanishes only modulo this gauge transformation.

Therefore eq (\ref{4.9}) is consistent and one can proceed to derive the
pure spinor action living with eq. (\ref{4.9}). This is done in subsection
4.1 following \cite{tor1}

However, even if (\ref{4.9}) is  consistent, one could
 be disappointed by the fact that the square of a nilpotent charge gives a
non vanishing result acting on some object. It is possible to avoid
this result by fixing the $\omega$-gauge symmetry and using the so
called Y-formalism \cite{Y-form1}, \cite{Y-form2} - \cite{Sken}. In
this formalism one can construct the pure spinor action in a way
which is more on line with the original proposal considered in
\cite{Oda1} for the heterotic string.This is done in subsection 4.2.
\subsection{Derivation of the pure spinor action}

The general method to construct pure spinor superstring actions in
generic backgrounds is the following. One starts from the
Green-Schwarz action  $I_{GS}$ in  conformal gauge, computes its BRS
variation and then adds a ``gauge fixing'' action $ I_{gf} $  given
by the (left-handed and right-handed) BRS variation of a suitable
``gauge fermion'' ( a functional with ghost number $n_{g} = - 1$)
such that the total action is BRS invariant.

$I_{gf}$ has to satisfy two conditions: \\

i) Its BRS variation must  cancel the variation of the Green-Schwarz term.\\

ii) It must be invariant under the gauge transformations
(\ref{4.10}) i.e. its dependence on $\omega_{R/L}$ should involve
only the terms  $(\omega_{R/L}
\lambda_{L/R})$, $(\omega_{R/L}\Gamma_{ab}\lambda_{L/R})$ and $(\omega_{R/L}
d \lambda_{L/R})$ . \\

The Green-Schwarz action is
\begin{eqnarray}
 I_{GS} = {1 \over 2}\int [ E_{+}^{a}\eta_{ab} E_{-}^{b} + B_{+ -}]
\label{4.13}
\end{eqnarray}
where  $B_{+ -}$ is the pullback of the NS-NS superform $B_{2}$.

The BRS variation of $I_{GS}$ is

\begin{eqnarray}
s I_{GS} = \int [ (\lambda_{L}E_{-}^{a}\Gamma_{a}E_{+ L}) +
   (\lambda_{R}E_{+}^{a}\Gamma_{a}E_{-R})]
\label{4.14}
\end{eqnarray}
A term which fulfils the condition i) is
\begin{eqnarray}
I^{(0)}_{gf} =  - \int s [ (E_{+ L}\omega_{R}) +  (E_{- R }\omega_{ L })]
\label{4.15}
\end{eqnarray}
Indeed, taking into account  that  $s^{2}$ is non vanishing only
when acts on $\omega_{L/R}$ and using (\ref{4.9}), one can see that
$  s I^{(0)}_{gf}$ cancels the variation of $I_{GS}$. However this
term cannot be  the whole story since when $ \tilde s_{R/L}$ acts on
$E_{\pm L/R}$,   the term
  $$ I^{(1)}_{gf}= - \int [ (\omega_{ R} M E_{+}^{a}\Gamma_{a}
\lambda_{R}) - (\lambda_{L} E_{-}^{a}\Gamma_{a}M\omega_{L})]$$ arises,
which is incompatible with the condition ii). Therefore, following \cite{tor1},
we replace $  I^{(0)}_{gf}$ with $$ I^{'(0)}_{gf}= I^{(0)}_{gf} -
I^{(1)}_{gf} =
 - \int [ s_{L} (E_{+ L}\omega_{R}) + s_{R} (E_{- R}\omega_{ L })]. $$
$I^{(1)}_{gf}$ can be rewritten as
\begin{eqnarray}
I^{(1)}_{gf} =  - \int (s_{R}^{2} - s_{L}^{2})(\omega_{R}M \omega_{ L})
\label{4.16}
\end{eqnarray}
This expression suggests to add the term
\begin{eqnarray}
I^{(2)}_{gf} =  1/2 \int (s_{L}s_{R} -s_{R}s_{L})(\omega_{R} M \omega_{
 L}) = \int s_{L}s_{R} (\omega_{R} M \omega_{ L})
\label{4.17}
\end{eqnarray}
in order to cancel the variation of $ I^{(1)}_{gf}$. As a result,
the ``gauge fixing'' action is
\begin{eqnarray}
I_{gf} = I^{(0)}_{gf} - I^{(1)}_{gf} + I^{(2)}_{gf} =  - \int [ s_{L} (
E_{+ L}\omega_{R}) + s_{R} (E_{- R}\omega_{ L })]
+  \int s_{L}s_{R} (\omega_{R} M \omega_{ L})
\label{4.18}
\end{eqnarray}
This expression for the ``gauge fixing'' action has been proposed in
\cite{tor1} and shown to be BRS invariant. In fact
the variation of  $I^{(0)}_{gf}$ cancels the variation of the G-S action, as
already noted, and $$ s I^{(1)}_{gf} = s I^{(2)}_{gf}. $$
Indeed  the variation of  $I^{(1)}_{gf}$ is
$$ s  I^{(1)}_{gf} = \int (s_{L}s_{R}^{2} - s_{R}s_{L}^{2})(\omega_{R}(
e^{\phi} P)\omega_{L})$$ and the variation of $I^{(2)}_{gf}$ is
$$ s  I^{(2)}_{gf} = \int (s_{L}^{2}s_{R} +s_{R}s_{L}s_{R})(\omega_{R}(
e^{\phi} P)\omega_{L})$$  Since $ (s_{L}s_{R} +s_{R}s_{L})$ always vanishes,
the r.h.s.'s of these two equations are equal.

Now let us compute $ I_{gf}$.

Using (\ref{3.12}), (\ref{3.13}) and (\ref{4.4}) one gets for  $ I^{(0)}_{gf}-
 I^{(1)}_{gf} $

\begin{eqnarray}
I^{(0 )}_{gf}- I^{(1)}_{gf} = \int [ (E_{+ L}d_{R}) +  (E_{- R }d_{L})
+ (\omega_{R}\Delta^{(L)}_{+}\lambda_{L}) +(\omega_{ L }\Delta^{(R)}_{-  }
\lambda_{R})]
\label{4.19}
\end{eqnarray}
where
\begin{eqnarray}
(\Delta^{(L/R)}\lambda_{L/R})^{\underline{\alpha}} = d \lambda^{
\underline{\alpha}} - \lambda^{\underline{\beta}}\Omega_{(L/R)\underline{\beta}
}{}^{ \underline{\alpha}} - {1 \over 4}
\lambda^{\underline{\alpha}}(E_{L/R}D_{L/R}\phi)
\label{4.20}
\end{eqnarray}
and $  \Omega_{(L/R)\underline{\beta}}{}^{\underline{\alpha}}$ is
defined as
\begin{eqnarray}
 \Omega_{(L/R) \underline{\beta}}{}^{\underline{\alpha}} = {1 \over 4}
(\Gamma_{ab})_{\underline{\beta}}{}^{\underline{\alpha}}( \Omega^{ab} +
{1 \over 2}(E_{L/R}\Gamma^{ab}
D_{L/R}\phi)\mp 3 E^{c}H_{c}{}^{ab})
\label{4.21}
\end{eqnarray}
Note that the second term  in the variation of $\omega_{L/R}$ (see
equations (\ref{4.4}), (\ref{4.1}) ) is essential for the
consistency of the result since it cancels the terms
$1/4[(\omega_{R/L} E_{L/R})(\lambda_{L/R}D_{L/R}
\phi) + 1/2 (\omega_{R/L}\Gamma^{ab}E_{L/R})(D_{L/R}\phi
\Gamma_{ab}\lambda_{L/R})]$ coming from the variation of $E_{L/R}$ (equations
(\ref{3.12}), (\ref{3.13})). In the absence of this term in the
r.h.s. of (\ref{4.4}) the result would be inconsistent with the
condition ii). This is an important consistency check of the ansatz
(\ref{4.1}) for $ X_{\underline{\beta}
\underline{\gamma}}{}^{\underline{\alpha}}$.

To compute $ I^{(2)}_{gf} $ it is convenient to use the first
expression for
 $I^{(2)}_{gf}$ in (\ref{4.17}) to  get
\begin{eqnarray}
 I^{(2)}_{gf} = - \int (\tilde s _{L}(\omega_{R}e^{\phi}) P e^{-\phi}
 \tilde s _{R}(\omega_{L}e^{\phi})) - \int  (\tilde s _{L}(\omega_{R}e^{\phi})
(\tilde s_{R} P) \omega_{L}) - \int (\omega_{R}(\tilde s_{L} P)
 \tilde s _{R}(\omega_{L}e^{\phi}))
\nonumber
\end{eqnarray}
\begin{eqnarray}
  + {1 \over 2} \int e^{\phi}(\omega_{R \alpha}((\tilde s_{R}\tilde s_{L}
 - \tilde s_{L}\tilde s_{R}) P^{\alpha \hat \alpha})\omega_{L \hat \alpha})
\nonumber
\end{eqnarray}
\begin{eqnarray}
  + {1 \over 2 }\int e^{\phi} [(\tilde s_{R}\tilde s_{L}\omega_{R
\alpha} + (\tilde s_{R}\tilde s_{L}\phi)\omega_{R \alpha})
P^{\alpha \hat \alpha}\omega_{L \hat \alpha} - \omega_{R \alpha}P^{\alpha
\hat \alpha}(\tilde s_{L}\tilde s_{R}\omega_{L \hat \alpha} +
(\tilde s_{L}\tilde s_{R}\phi)\omega_{L \hat \alpha})]
\label{4.22}
\end{eqnarray}
The terms involving $ \tilde s_{R/L}\tilde s_{L/R}\phi$ in the last integral
arise to avoid double counting.

The first three terms in the r.h.s. of (\ref{4.22}) are consistent with the
condition ii) since the monomials $\lambda^{\underline{\alpha}}\omega_{
\underline{\beta}}$ are saturated by LW-valued superfields. In fact
\begin{eqnarray}
I^{(a)}_{gf} \equiv \int (\tilde s_{L}(\omega_{R}e^{\phi}) P e^{\phi}
 \tilde s_{R}(\omega_{L}e^{\phi})) = \int  (d_{R \alpha} - \lambda_{L}{}^{
\beta}\omega_{R{}\gamma}Y_{\beta \alpha}{}^{\gamma})e^{\phi}
P^{\alpha \hat \alpha}(d_{L \hat  \alpha} - \lambda_{R}{}^{\hat \beta}
\omega_{L{} \hat \gamma}Y_{\hat \beta \hat \alpha}{}^{\hat \gamma})
\label{4.23}
\end{eqnarray}
\begin{eqnarray}
 I^{(b)}_{gf} \equiv \int  (\tilde s _{L}(\omega_{R}e^{\phi})(\tilde s_{R} P)
\omega_{L}) = -
  (d_{R \alpha} - \lambda_{L}{}^{\beta}
\omega_{R \gamma}Y_{\beta \alpha}{}^{\gamma })e^{\phi}
C_{R \hat \beta}{}^{\alpha \hat \gamma}\lambda_{R}{}^{\hat \beta}\omega_{L
\hat \gamma}
\label{4.24}
\end{eqnarray}
and
\begin{eqnarray}
 I^{(c)}_{gf} \equiv \int (\omega_{R}(\tilde s_{L} P)\tilde s _{R}
(\omega_{L}e^{\phi}))= -
 \lambda_{L}{}^{\beta}\omega_{R \alpha}e^{\phi}C_{L{}\beta}{}^{\alpha
\hat \alpha}  (d_{L{} \hat  \alpha} - \lambda_{R}{}^{\hat \beta}
\omega_{L \hat \gamma} Y_{\hat \beta \hat \alpha}{}^{\hat \gamma })
\label{4.25}
\end{eqnarray}
The last two integrals in (\ref{4.22}), that is,
\begin{eqnarray}
I^{(d)}_{gf} =  {1 \over 2} \int e^{\phi}(\omega_{R \alpha}
((\tilde s_{R}\tilde s_{L} - \tilde s_{L}\tilde s_{R}) P^{\alpha
\hat \alpha})\omega_{L \hat \alpha})
\label{4.250}
\end{eqnarray}
 and
\begin{eqnarray}
 I^{(e)}_{gf} =
 {1 \over 2 }\int e^{\phi} [(\tilde s_{R}\tilde s_{L}\omega_{R
\alpha} + (\tilde s_{R}\tilde s_{L}\phi)\omega_{R \alpha})
P^{\alpha \hat \alpha}\omega_{L \hat \alpha} - \omega_{R \alpha}P^{\alpha
\hat \alpha}(\tilde s_{L}\tilde s_{R}\omega_{L \hat \alpha} +
(\tilde s_{L}\tilde s_{R}\phi)\omega_{L \hat \alpha})]
\label{4.251}
\end{eqnarray}
 are potentially dangerous since both give
contributions that violate condition  ii). Luckily these contributions  cancel
 each other. This is proved in the Appendix where the LW-valued
contributions of the last two integrals are computed. The result is
\begin{eqnarray}
I^{(d)}_{gf} + I^{(e)}_{gf} =  \int e^{\phi}\omega_{R \beta}\lambda_{L}
{}^{\alpha}\omega_{L \hat \gamma}\lambda_{R}{}^{\hat \delta} C_{ \alpha
\hat \delta}{}^{\beta \hat \gamma} +
 \int e^{\phi}\omega_{R \beta}\lambda_{L}{}^{\alpha}\omega_{L
\hat \gamma}\lambda_{R}{}^{\hat \delta} \tilde C_{ \alpha \hat \delta}
{}^{\beta \hat \gamma}
\label{4.27}
\end{eqnarray}
where
\begin{eqnarray}
C_{\alpha \hat \delta}{}^{\beta \hat \gamma} =  \Pi_{\alpha \sigma}
{}^{\beta \tau}(\Delta_{[\tau}\Delta_{\hat \tau ]}  P^{\sigma \hat \sigma})
\Pi_{\hat \sigma \hat \delta}{}^{\hat \tau \hat \gamma}
\label{4.30}
\end{eqnarray}
and
\begin{eqnarray}
\tilde C_{ \beta \hat \delta}{}^{\gamma \hat \gamma} = e^{\phi}
 \Pi_{\beta \tau}{}^{\gamma \sigma} [R_{\sigma \hat \sigma \rho}{}^{\tau}
P^{\rho \hat \tau} - R_{\sigma \hat \sigma \hat \rho}{}^{\hat \tau}
P^{\tau \hat \rho} + (\Gamma^{c} P \Gamma_{c})_{\sigma \hat \sigma}
P^{\tau \hat \tau}]
\Pi_{ \hat \tau \hat \delta}{}^{\hat \sigma \hat \gamma}
\label{4.31}
\end{eqnarray}
where $$  \Pi_{\alpha \sigma}{}^{\beta \tau} = \delta_{\alpha}{}^{\beta}
\delta_{\sigma}{}^{\tau} - {1 \over {16.4!}}(\Gamma^{abcd})_{\alpha}{}^{\beta}
(\Gamma_{abcd})_{\sigma}{}^{\tau} $$ is the projector that projects on the
LW-valued component of a superfield $Y_{\tau}{}^{\sigma}$    and
$ \Pi_{\hat \sigma \hat \delta}{}^{\hat \tau \hat \gamma}$  is  defined
in a similar way in terms of the hatted quantities.

Adding to the Green-Schwarz action (\ref{4.13}) the contributions of
$ I_{gf}$ which are given in  (\ref{4.19}), (\ref{4.23}),
(\ref{4.24}), (\ref{4.25}), and (\ref{4.27}) one obtains  the pure
spinor  action for type IIA superstring in a general background:
\begin{eqnarray}
I = \int [ {1 \over 2}(E_{+}^{a}E_{- a} + B_{ + -}) + E_{+ L}^{
\alpha} d_{R \alpha} +  E_{- R}^{\hat \alpha} d_{L \hat \alpha}
\nonumber
\end{eqnarray}
\begin{eqnarray}
+ \omega_{R \alpha}(\Delta_{+}^{(L)}\lambda_{L})^{\alpha} +  \omega_{L
\hat \alpha }(\Delta_{-}^{(R)} \lambda_{R})^{\hat \alpha}
- d_{R \alpha}M^{\alpha \hat \beta}d_{L \hat \beta}
\nonumber
\end{eqnarray}
\begin{eqnarray}
 + d_{R \gamma}e^{\phi}\tilde C^{\gamma \hat \alpha}{}_{\hat \beta} \omega_{L
\hat \alpha}
\lambda_{R}{}^{\hat \beta}  + \omega_{R \beta}\lambda_{L}^{\alpha}e^{\phi}
 \tilde C^{\hat \gamma \beta}{}_{\alpha}
d_{L \hat \gamma}  -  \omega_{R \alpha}\lambda_{L}^{\beta}
S^{\alpha \hat \gamma}{}_{\beta \hat \delta}\omega_{L \hat \gamma}
\lambda_{R}^{\hat \delta}]
\label{4.32}
\end{eqnarray}
where  $\Delta^{(L/R)}_{\pm}$ are the pullbacks of the covariant
differentials defined in (\ref{4.20}), (\ref{4.21}). $ \tilde
C^{\gamma \hat \alpha}{}_{\hat
\beta}$,   $\tilde C^{\hat \gamma \alpha}{}_{\beta}$ and
$S^{\alpha \hat \gamma}_{\beta \hat \delta}$  are given by
$$ \tilde C^{\gamma \hat \alpha}{}_{\hat
\beta} =  C_{R \hat \beta}{}^{\hat \alpha \gamma} + P^{\gamma \hat \gamma}
Y_{\hat \beta \hat \gamma}{}^{\hat \alpha}  $$   $$ \tilde C^{\hat \gamma
\alpha}{}_{\beta} =  C_{L \beta}{}^{\alpha \hat \gamma} + P^{\gamma
\hat \gamma}Y_{ \beta \gamma}{}^{\alpha}   $$
 $$ S^{\alpha \hat \gamma}_{\beta \hat \delta} =  C^{\alpha \hat \gamma}_{
\beta \hat \delta} + \tilde C^{\alpha \hat \gamma}_{\beta \hat \delta}  +
Y_{\beta \gamma}{}^{\alpha}P^{\gamma \hat \beta} Y_{\hat \delta \hat \beta}{}
^{\hat \gamma} $$
where $P$, $C_{R \hat \beta}{}^{\hat \alpha \gamma}$, $ C_{L \beta}{}^{
\alpha \hat \gamma}$, $Y_{\hat \beta \hat \gamma}{}^{\hat \alpha}$,
$ Y_{\beta \gamma}{}^{\alpha}$, $   C^{\alpha \hat \gamma}_{\beta
\hat \delta}$ and $ \tilde C^{\alpha
\hat \gamma}_{\beta \hat \delta}$ are defined in the equations (\ref{2.22}),
(\ref{3.24}), (\ref{4.12}), (\ref{4.30}), (\ref{4.31}) so that all
the superfields in (\ref{4.32}) are given explicitly in terms of the
components of torsions and curvatures, or more specifically in terms
of $ P $,
 $\phi$ and their (covariant) derivatives.

The action  is manifestly invariant under BRS transformations as well as under
the  gauge tranformation (\ref{4.10}) of $\omega$ . Moreover the field
equations obtained from the action (\ref{4.32})  varying
$d_{\underline{\alpha}}$ are
\begin{eqnarray}
 E^{\alpha}_{+ L} = - e^{\phi}( P_{L}^{\alpha \hat \beta}d_{L \hat \beta} -
\tilde C^{ \alpha \hat \beta}{}_{\hat \gamma}\lambda_{R}^{\hat \gamma}
\omega_{L \hat \beta} )
\nonumber
\end{eqnarray}
\begin{eqnarray}
 E^{\hat \alpha}_{- R} =  e^{\phi}(
P_{R}^{\hat \alpha  \beta}d_{R  \beta} -
\tilde C^{\hat \alpha  \beta}{}_{ \gamma}\lambda_{L}^{\gamma}
\omega_{R \beta} )\,.
\label{4.33}
\end{eqnarray}
They are identical to equations (\ref{4.8}) and assure the on shell
nilpotency of $ s $ acting on $ d_{\underline{\alpha}}$. The action
(\ref{4.32}) is precisely the action first obtained in \cite{BH}.

\subsection{Alternative derivation of the  action. The
Y-formalism}

{\underline{Y-formalism}}.

If the equal time Poisson Brackets (ETPB's)  among
$\omega $ and $ \lambda $ are the canonical ones,  the ETPB's
among $\omega_{R/L}$ and $(\lambda_{L/R}\Gamma^{a}\lambda_{L/R})$ do not
vanish:
\begin{eqnarray}
\lbrace \omega(\sigma)_{R/L \underline{\alpha}}, (\lambda(\sigma')_{L/R}
\Gamma^{a}\lambda(\sigma')_{L/R}) \rbrace =  2
(\Gamma^{a}\lambda_{L/R}(\sigma'))_{\underline{\alpha}}
 \delta(\sigma' - \sigma)
\label{5.1}
\end{eqnarray}
 The constraint $ (\lambda_{L/R}\Gamma^{a}\lambda_{L/R}) = 0 $
generates the gauge transformation (\ref{4.10}) of $\omega_{R/L}$,
but in the pure spinor approach this constraint is assumed to hold in a
strong sense and therefore equation (\ref{5.1})  is unsatisfactory. This
problem  can be avoided \cite{Ber1}, \cite{Oda1} by assuming the following
ETPB among  $\omega $ and $ \lambda $:
\begin{eqnarray}
\lbrace \omega_{ \underline{\alpha}}( \sigma), \lambda^{\underline{\beta}}(
\sigma')\rbrace = \delta(\sigma - \sigma' )[ \delta_{\underline{\alpha}}{}^{
\underline{\beta}} - K_{\underline{\alpha}}{}^{\underline{\beta}}]
\label{5.2}
\end{eqnarray}
where $ K_{\underline{\alpha}}{}^{\underline{\beta}} \equiv ( K_{L \alpha}{}^{
\beta}, K_{R \hat \alpha}{}^{\hat \beta}) $ are the projectors
\begin{eqnarray}
K_{L \alpha}{}^{\beta} = {1 \over 2} (\Gamma^{a}\lambda_{L})_{\alpha}
(Y_{R}\Gamma_{a})^{\beta}
\nonumber
\end{eqnarray}
\begin{eqnarray}
K_{R \hat \alpha}{}^{\hat \beta} = {1 \over 2} (\Gamma^{a}\lambda_{R})_{
\hat \alpha}
(Y_{L}\Gamma_{a})^{\hat \beta}
\label{5.3}
\end{eqnarray}
with $$ Y_{R/L \underline{\alpha}} = { {V_{R/L \underline{\alpha}}} \over
{(V_{R/L} \lambda_{L/R}})} $$    so that $$  (Y_{R/L}\lambda_{L/R}) = 1 .$$
 If one chooses  $ V_{R/L} $  constant, $ K_{\underline{\alpha}}{}^{
\underline{\beta}} $ breaks  Lorentz
invariance. Moreover it is singular at $ (V_{L/R} \lambda_{L/R}) = 0 $.

In the case of IIA
superstrings one can avoid the breaking of Lorentz invariance by choosing
\begin{eqnarray}
 Y_{R/L} = { {\lambda_{R/L}} \over {(\lambda_{R}\lambda_{L})}}.
\label{5.4}
\end{eqnarray}
In the following we shall adopt this choice. With this choice, if one defines
$$ K^{\alpha}{}_{\beta} =  {1 \over 2} (\Gamma^{a}\lambda_{R})^{\alpha}{
1 \over {(\lambda_{L}\lambda_{R})}}(\lambda_{L}\Gamma_{a})_{\beta},
$$
$K_{R}$ and $K_{L}$ are transposed to each other, and
$$ K^{\alpha}{}_{\beta} =  (K_{R})^{\alpha}{}_{\beta}. $$
Even if now
the Lorentz invariance is preserved, $K$ is still singular when
$(\lambda_{R}\lambda_{L}) = 0 $. In any case these deseases - breaking of
Lorentz invariance and/or singular nature of $K$ - are innocuous since, as we
shall see, any dependence of $K$ will be absent in the final results.

 {\underline{Gauge fixing}}.

One can gauge fix  the $\omega$-gauge symmetry (\ref{4.10}) by
requiring  $$ (\omega_{R/L}\Gamma^a \lambda_{R/L})=0 .$$
Using the projector $ K $ , this gauge fixing condition is
equivalent to
\begin{eqnarray}
(K\omega_{L}) = 0 = (\omega_{R} K)
\label{5.51}
\end{eqnarray}
or
\begin{eqnarray}
\omega_{R \alpha} = \omega_{R \beta}(1 - K)^{\beta}{}_{\alpha} \qquad
\omega_{L}{}^{\alpha} = (1-K)^{\alpha}{}_{\beta}\omega_{L}{}^{\beta}
\label{5.5}
\end{eqnarray}
which are consistent with (\ref{5.2}).
Also notice that
\begin{eqnarray}
( K \lambda_{L})^{\alpha} = 0 = (\lambda_{R} K )_{\alpha}
\label{5.6}
\end{eqnarray}
and
\begin{eqnarray}
(\lambda_{L}\Gamma^{a} (1 - K))_{\alpha} = 0 =
((1 - K)\Gamma^{a}\lambda_{R}){}^{\alpha}
\label{5.7}
\end{eqnarray}
Equation (\ref{5.7}), toghether with (\ref{5.5}) implies that
$\omega_{R/L}$ have vanishing ETPB's with the
constraint (\ref{3.9}).  Moreover, since $Tr K_{R/L} = 5 $,  $K_{L}$ and
$K_{R}$ project on
five-dimensional subspaces of the 16-dimensional spinorial spaces and
therefore each of the ghosts $ \lambda_{L}$, $\lambda_{R}$, $\omega_{R}$ and
$\omega_{L}$ has eleven components.
The fields $d_{\underline{\alpha}}$ can be splitted as
$$ d^{\top}_{R \alpha} = (d_{R}(1 - K))_{\alpha}$$
$$ d_{L}^{\top \alpha} = ((1 - K) d_{L})^{\alpha}$$
$$ d^{\bot}_{R \alpha} = (d_{R} K)_{\alpha}$$
$$ d_{L}^{\bot \alpha} = (K d_{L})^{\alpha}.$$

 Only $ d_{R/L}^{\top} $ appears  in the BRS charge $Q$ so that $
d_{R/L}^{\top}$ are the BRST partners of $\omega_{R/L}$. With these
definitions, the BRST transformations of $\omega_{
\underline{\alpha}}$, $ d^{\top}_{\underline{\alpha}}$,  $ d^{\bot}_{
\underline{\alpha}}$ can be obtained by projecting (\ref{4.4}) and (\ref{4.5})
on the subspaces spanneds by the projectors $ K $ and $(1 - K)$.
In particular
\begin{eqnarray}
\tilde s d^{\bot}_{R \alpha}= - [(\lambda_{L}E_{-}^{a}
\Gamma_{a})_{\gamma} - 2 (\lambda_{L}\lambda_{R}) (\omega_{R}M)_{\gamma} +
(\lambda_{L}D_{R}\phi)
d^{\bot}_{R \gamma} - \lambda_{L}^{\beta}Y_{\beta \gamma}{}^{\delta}d^{\bot}{}
_{ R \delta}] K^{\gamma}{}_{\alpha}
\nonumber
\end{eqnarray}
\begin{eqnarray}
\tilde s d_{L}^{\bot \alpha} = -   K^{\alpha}{}_{
\gamma}[(E_{+}^{a}\Gamma_{a}\lambda_{R})^{\gamma} + 2 (\lambda_{L}\lambda_{R})
(M\omega_{L})^{\gamma} +
(\lambda_{R} D_{L}\phi)d^{\bot \gamma} - \lambda_{R}^{\hat \beta}Y_{\hat \beta
\hat \tau}{}^{\hat \sigma}d^{\bot}_{L \hat \sigma}\delta^{\hat \tau \gamma}]
\label{5.10}
\end{eqnarray}

 Projecting (\ref{4.7}), (\ref{4.9}) with $(1 - K)$, one has
\begin{eqnarray}
s^{2} d^{\top}_{R/L} = 0 = s^{2} \omega_{R/L}
\label{5.11}
\end{eqnarray}
Moreover
\begin{eqnarray}
s^{2} d^{\bot}_{R \alpha} =  - (\lambda_{R}\lambda_{L})
(E_{-  R} + \tilde s_{L}(\omega_{R}M))_{\gamma} K^{\gamma}{}_{\alpha}
\nonumber
\end{eqnarray}
\begin{eqnarray}
s^{2} d^{\bot \alpha}_{L} = - (\lambda_{R}\lambda_{L})K^{\alpha}{}_{
\gamma} (E_{+ L} - \tilde s_{R}(M\omega_{L}))^{\gamma}
\label{5.12}
\end{eqnarray}
The right hand sides of (\ref{5.12}) vanish on shell if  $E_{\mp
R/L}$ satisfy the field equations (\ref{4.8}). It follows from
(\ref{5.11}), (\ref{5.12}) that now s is nilpotent acting on any
field or ghost.

{\underline{Derivation of the Action}}

In  this formalism the strategy  to derive the pure spinor action is
similar to
\cite{Oda1}. Add to the Green-Schwarz action a new $ K $-dependent
term $ I_{(K)}$ such that $ I_{GS} +  I_{(K)} $ is
 BRST invariant. Then add the ``gauge fixing''  term $ I_{gf} =  s \int
{ F } $ to cancel  the dependence on $K$ in  the total action. The  ``gauge
fermion'' $ F$ is a local functional with $ n_{gh}= - 1 $.
 Since $ s $ is always nilpotent, $ I_{gf}$ is automatically BRS
invariant.

A possible choice of $I_{(K)}$ is
\begin{eqnarray}
I_{(K)} = - \int [ (d_{R} K E_{+ L}) - (E_{-R} K d_{L})] - \int (d_{R} K M K
d_{L}) - 2 \int (\lambda_{L}\lambda_{R})(\omega_{R}M_{R}KM_{L}\omega_{L})
\label{5.13}
\end{eqnarray}
Indeed
$$ \tilde s [(d_{R} K E_{+ L}) - (E_{- R} K d_{L})] =[(\lambda_{L}E_{-}^{a}
\Gamma_{a}E_{+ L}) + (\lambda_{R}E_{+}^{a}\Gamma_{a}E_{- R})] - 2
(\lambda_{L}\lambda_{R})[ (\omega_{R} M K E_{+L})  -(E_{- R} K M \omega_{L})]
 $$
$$ + [ (d_{R} K M E_{+}^{a}\Gamma_{a}\lambda_{R}) - (\lambda_{L}\Gamma_{a}
E_{-}^{a} M K d_{L})] $$

$$ \tilde s (d_{R} K M K d_{L}) = - [ (d_{R} K M E_{+}^{a}\Gamma_{a}
\lambda_{L}) - (\lambda_{R}\Gamma_{a}E_{-}^{a}M K d_{L})] -
 2 (\lambda_{L}\lambda_{R})[ (\omega_{R} M K M K d_{L}) + (d_{R}K M  K M
\omega_{L})] $$ and $$ 2 \tilde s [ (\lambda_{L}\lambda_{R}) (\omega_{R} M K M
\omega_{L})] = - 2 (\lambda_{L}\lambda_{R})[ (\omega_{R} M K M (1 - K) d_{L}) +
 (d_{R}(1 - K) M  K M \omega_{L})]$$ $$ + 2 (\lambda_{L}\lambda_{R})e^{\phi} [
\omega_{R}(\lambda_{L}^{\alpha}\tilde C_{L \alpha}) K M \omega_{L}) + (\omega_
{R} M K (\lambda_{R}^{\hat \alpha}\tilde C_{R \hat \alpha})\omega_{L})] $$
so that $$ s I_{GF} + sI_{(K)} = 0. $$ modulo the field equations (\ref{4.33}).

Then choosing
\begin{eqnarray}
I_{gf} = s \int [ (d_{R} K M \omega_{L}) - (\omega_{R}  M K d_{L})] -
s \int [(\omega_{R} E_{+ L}) + (E_{-R}\omega_{L})]
\nonumber
\end{eqnarray}
\begin{eqnarray}
+ {1 \over 2}s \int [(\tilde s_{R} -
\tilde s_{L}) (\omega_{R} M \omega_{L})]
\label{5.14}
\end{eqnarray}
one can verify that the total action $$ I = I_{GS} + I_{(K)} + I_{gf}$$
reproduces the pure spinor action (\ref{4.32}).

\section{Conclusion}

To summarize,  in this paper, generalizing the method of Extended
Differential Algebra, proposed in \cite{tor1}, we have shown that it
is possible to start from a geometrical formulation of Type IIA
supergravity (rheonomic parametrization of torsions and curvatures)
and derive the standard pure spinor constraints, the nilpotency of
the BRST charge, and the BRST invariant action of the pure spinor
superstrings formulation.The pure spinor constraints follow from the
requirement that the ghosts $\lambda^{a}$, related to the
vector-like supervielbeins, vanish (similar to the superembedding
constraint) together with the mild assumption that
 $ (\lambda_{L}\lambda_{R}) $ does not generically vanish.

In a sense this reverses the pattern followed in \cite{BH} where,
starting from the nilpotency of the BRST charge and the most general
BRST invariant action (or, equivalently, the holomorphicity
properties of the BRST currents), a consistent set of  on shell
supergravity constraints is derived.

The results that we obtain are equivalent to those of \cite{BH}
modulo the different choice of the supergravity constraints.
However, the supergravity constraints from which we start (and the
superfields which appear in the final action) differ from those
derived in
\cite{BH} at most by a redefinition of supervielbeins, superconnections,
gauge superforms and the dilaton. The fact that in \cite{BH} the
structure group is very large (it involves three independent  local
Lorentz groups, for vector, left-handed  and right-handed spinors
and two independent local Weyl groups for left-handed and
right-handed spinors) should not deceive. Indeed, as shown in
\cite{BH},  this large gauge symmetry must be gauge fixed and
reduced to a single local Lorentz invariance in order  to cancel
some spurious superfields in torsion components.  However the form of the 
left-handed and
right-handed Lorentz and (gauge fixed) Weyl connections
remain different and this is also an a posteriori result of our
approach ( see (\ref{4.20}) and (\ref{4.21})).

The pure spinor action has been derived in two ways. In particular
the second derivation is a generalization to the case of IIA
superstring of a procedure first proposed in \cite{Oda1} for the
heterotic string. An advantage of this method is that, once
the Green-Schwarz action is modified by the addition of suitable
K-dependent terms in order to promote its k-symmetry to a
pure-spinor BRS symmetry, the remaining step to get the pure spinor
action is a standard BRS-like gauge fixing procedure i.e. the
addition of a BRS exact , local action.

The rheonomic parametrization that we have adopted in this paper is  that
considered in \cite{tor1}. However, since all the consistent parametrizations
are equivalent (in the sense specified above) it is quite evident that the
procedure described in this paper can be used starting with any consistent
rheonomic parametrization.

\section{Acknowledgments}

I would like to thank D. Sorokin for valuable discussions and advices and I.
Bandos and I. Oda for useful comments.

\medskip
\medskip
\def\theequation{A.\arabic{equation}}
\def\thesection{Appendix.}
\section{Computation of $ I_{gf}^{(d)} + I_{gf}^{(e)} $ }
\setcounter{equation}0

In this Appendix we study the integrals $I^{(d)}_{gf}$ and $I^{(e)}_{gf}$
defined in (\ref{4.250}) and (\ref{4.251}) and show that the contributions of
these integrals that violate the $\omega$-gauge invariance cancel each other.

To compute $I^{(d)}_{gf}$ let us  consider the left-handed and right-handed
BRS variations of eqs.(\ref{3.24})
\begin{eqnarray}
\tilde s_{R} \tilde s_{L}P^{\beta \hat\gamma} \equiv \lambda_{L}{}^{\alpha}
\lambda_{R}{}^{\hat \delta}\Delta_{\hat \delta}C_{L{} \alpha}{}^{\beta
\hat \gamma}
\label{A.1}
\end{eqnarray}
\begin{eqnarray}
\tilde s_{L} \tilde s_{R}P^{\beta \hat\gamma} \equiv \lambda_{L}{}^{\alpha}
\lambda_{R}{}^{\hat \delta}\Delta_{\alpha}C_{R{} \hat \delta}{}^{\beta
\hat \gamma}
\label{A.2}
\end{eqnarray}
It follows from (\ref{3.25}) that  $ \Delta_{\hat \delta}C_{L{} \alpha}{}^{
\beta \hat \gamma}$ and $ \Delta_{\alpha}C_{R{} \hat \delta}{}^{\beta
\hat \gamma}$ are LW-valued with respect to $\alpha, \beta $ and $\hat \delta,
\hat \gamma $ respectively, so that  we can write
\begin{eqnarray}
{1 \over 2} \int e^{\phi}\omega_{R \alpha}((\tilde s_{R}\tilde s_{L}
 P^{\alpha \hat \alpha})\omega_{L \hat \alpha})
=  \int e^{\phi}\omega_{R \beta}\lambda_{L}{}^{\alpha}\omega_{L \hat \gamma}
\lambda_{R}{}^{\hat \delta}[ C_{L \alpha \hat \delta}
{}^{\beta \hat \gamma} +
\Xi_{L \alpha \hat \delta}{}^{\beta \hat \gamma}]
\label {A.3}
\end{eqnarray}
and
\begin{eqnarray}
{1 \over 2} \int e^{\phi}\omega_{R \alpha}( \tilde s_{L}\tilde s_{R} P^{
\alpha \hat \alpha})\omega_{L \hat \alpha})
=  \int e^{\phi}\omega_{R \beta}\lambda_{L}{}^{\alpha}\omega_{L \hat \gamma}
\lambda_{R}{}^{\hat \delta}[ C_{R \alpha \hat \delta}
{}^{\beta \hat \gamma} + \Xi_{R \alpha \hat \delta}
{}^{\beta \hat \gamma}]
\label {A.4}
\end{eqnarray}
where $ C_{L/R \alpha \hat \delta}{}^{\beta \hat \gamma}$ are
LW-valued both in the indices $\alpha, \beta$ and $\hat \delta, \hat
\gamma$ whereas $ \Xi_{L \alpha \hat \delta}{}^{\beta \hat \gamma}$
is the contribution which is LW-valued in $\alpha, \beta$ but not in
$\hat
\delta, \hat \gamma$, the latter being proportional to  $ (\Gamma^{abcd})_{\hat
\delta}{}^{\hat \gamma}$. On the other hand $ \Xi_{R \alpha \hat \delta}{}^{\beta
\hat \gamma}$ is LW-valued in $\hat \delta, \hat
\gamma$ but not in $\alpha, \beta$ which is proportional to  $(\Gamma^{abcd})_{\alpha}{}^{\beta}$.

Therefore
\begin{eqnarray}
I^{(d)}_{gf} =
{1 \over 2} \int e^{\phi}\omega_{R \alpha}((\tilde s_{R}\tilde s_{L}
 - \tilde s_{L}\tilde s_{R}) P^{\alpha \hat \alpha})\omega_{L \hat \alpha}
\nonumber
\end{eqnarray}
\begin{eqnarray}
=  \int e^{\phi}\omega_{R \beta}\lambda_{L}{}^{\alpha}\omega_{L \hat \gamma}
\lambda_{R}{}^{\hat \delta}[ C_{ \alpha \hat \delta}
{}^{\beta \hat \gamma} +
(\Xi_{R \alpha \hat \delta}{}^{\beta \hat \gamma} - \Xi_{L \alpha \hat \delta}
{}^{\beta \hat \gamma})]
\label {A.5}
\end{eqnarray}
where
\begin{eqnarray}
C_{\alpha \hat \delta}{}^{\beta \hat \gamma} \equiv
 C_{L \alpha \hat \delta}{}^{\beta \hat \gamma} -  C_{R \alpha \hat \delta}
{}^{\beta \hat \gamma} =  \Pi_{\alpha \sigma}
{}^{\beta \tau}(\Delta_{[\tau}\Delta_{\hat \tau ]}  P^{\sigma \hat \sigma})
\Pi_{\hat \sigma \hat \delta}{}^{\hat \tau \hat \gamma}
\label{A.6}
\end{eqnarray}
 $  \Pi_{\alpha \sigma}{}^{\beta \tau}$ and $ \Pi_{\hat \sigma
\hat \delta}{}^{\hat \tau \hat \gamma}$  being  the projectors that project on
the well-behaved components of superfields $Y_{\alpha}{}^{\beta}$ and
$ Y_{\hat \delta}{}^{\hat \gamma} $ respectively.
 $  C_{ \alpha \hat \delta}{}^{\beta \hat \gamma}$ are LW-valued both in
 $\alpha, \beta$ and $\hat \delta, \hat \gamma$.

To compute $\Xi_{R/L} $  let us consider, instead of (\ref{A.5}), the integral
\begin{eqnarray}
{1 \over 2} \int e^{\phi}\omega_{R \alpha}((\tilde s_{R}\tilde s_{L}
  + \tilde s_{L}\tilde s_{R}) P^{\alpha \hat \alpha})\omega_{L \hat \alpha}
=  \int e^{\phi}\omega_{R \beta}\lambda_{L}{}^{\alpha}\omega_{L \hat \gamma}
\lambda_{R}{}^{\hat \delta}[ C_{L \alpha \hat \delta}
{}^{\beta \hat \gamma} +   C_{ R \alpha \hat \delta}{}^{\beta \hat \gamma}
\nonumber
\end{eqnarray}
\begin{eqnarray}
 +(\Xi_{R \alpha \hat \delta}{}^{\beta \hat \gamma} + \Xi_{L \alpha
\hat \delta}{}^{\beta \hat \gamma})]  =  -  \int e^{\phi}\omega_{R \beta}
\lambda_{L}{}^{\alpha}\omega_{L \hat \gamma}\lambda_{R}{}^{\hat \delta}
 [R_{\alpha \hat \delta \tau}{}^{\beta} P^{\tau \hat \gamma} + R_{\alpha
\hat \delta \hat \sigma}{}^{\hat \gamma}P^{\beta \hat \sigma}]
\label {A.7}
\end{eqnarray}
where the last equality follows taking into account the action of
$\tilde s^2 $ on Lorentz valued fields (see (\ref{3.09})).
By performing in (\ref{A.7}) a gauge transformation of $\omega_{L \hat \gamma}$
i.e.by replacing $\omega_{L \hat \gamma}$ with $ \Lambda_{c}(\lambda_{R}
\Gamma^{c})_{\hat \gamma}$ only   $\Xi_{L} $  survives and is determined
unambiguosly from the last identity in (\ref{A.7} ). The same can be
 repeated  for  $\Xi_{R} $  by replaciung  $\omega_{R \beta}$
with  $ \Lambda_{c}(\lambda_{L}\Gamma^{c})_{\beta}$.

Then to compute $I^{(e)}$ let us write $$ I^{(e)}_{gf} =  I^{(e 1)}_{gf} -
I^{(e 2)}_{gf} $$ where
\begin{eqnarray}
 I^{(e 1)}_{gf} =  {1 \over 2 }\int e^{\phi} (\tilde s_{R}\tilde s_{L}\omega_{R
\alpha} + (\tilde s_{R}\tilde s_{L}\phi)\omega_{R \alpha})
P^{\alpha \hat \alpha}\omega_{L \hat \alpha} \equiv \lambda_{L}{}^{\beta}
\omega_{R{} \gamma}\lambda_{R}{}^{\hat \delta}\omega_{L{} \hat \gamma}
  X_{(L) \beta \hat \delta}{}^{\gamma \hat \gamma}
\nonumber
\end{eqnarray}
\begin{eqnarray}
 I^{(e 2)}_{gf} =
{1 \over 2} \int e^{\phi}( \omega_{R \alpha}P^{\alpha
\hat \alpha}(\tilde s_{L}\tilde s_{R}\omega_{L \hat \alpha} +
(\tilde s_{L}\tilde s_{R}\phi)\omega_{L \hat \alpha})) \equiv \lambda_{L}{}^{
\beta}\omega_{R{} \gamma}\lambda_{R}{}^{\hat \delta}\omega_{L{} \hat \gamma}
  X_{(R) \beta \hat \delta}{}^{\gamma \hat \gamma}
\label{A.8}
\end{eqnarray}
 It follows from (\ref{4.11}) that  $X_{(L)}$ is
LW-valued in $\beta, \gamma $ and $ X_{(R)}$ is LW-valued  in $\hat \delta,
\hat \gamma$ so that we can write
$$   X_{L/R \beta \hat \delta}{}^{\gamma \hat \gamma} = \tilde C_{L/R \beta
\hat \delta}{}^{\gamma \hat \gamma} +
\Psi_{L/R \beta \hat \delta}{}^{\gamma \hat \gamma} $$
and
\begin{eqnarray}
I^{(e)}_{gf}= I^{(e 1)}_{gf} - I^{(e 2)}_{gf} = \lambda_{L}{}^{\beta}\omega_{
R{} \gamma}
\lambda_{R}{}^{\hat \delta}\omega_{L{} \hat \gamma}
(\tilde C_{ \beta \hat \delta}{}^{\gamma \hat \gamma} +
\Psi_{R \beta \hat \delta}{}^{\gamma \hat \gamma} -  \Psi_{L \beta
\hat \delta}{}^{\gamma \hat \gamma})
\label{A.9}
\end{eqnarray}
where $\tilde C \equiv \tilde C_{L} - \tilde C_{R} $ is LW-valued both in
  $\beta, \gamma$ and in
$\hat \delta, \hat \gamma$ and  $\Psi_{L \beta \hat
\delta}{}^{\gamma \hat \gamma}$
(  $\Psi_{R \beta \hat \delta}{}^{\gamma \hat \gamma}$) is LW-valued in
 $(\beta, \gamma)$  but not
in $(\hat \delta, \hat \gamma )$ ( in $ (\hat \delta,
\hat \gamma)$  but not in $(\beta, \gamma)$ ).
Since $ \tilde s_{R}\omega_{R} = 0 = \tilde s_{L}\omega_{L}$,
 $I^{(e 1)}_{gf}$ and  $I^{(e 2)}_{gf}$ can be rewritten as
\begin{eqnarray}
 I^{(e 1)}_{gf} = {1 \over 2 }\int e^{\phi} [(\tilde s_{R}\tilde s_{L} +
\tilde s_{L} \tilde s_{R})\omega_{R\alpha} +
(\tilde s_{R}\tilde s_{L}\phi)\omega_{R \alpha}]
P^{\alpha \hat \alpha}\omega_{L \hat \alpha }
\nonumber
\end{eqnarray}
\begin{eqnarray}
=  \int e^{\phi} [\lambda_{L}{}^{\alpha}
\omega_{R{} \beta}\lambda_{R}{}^{\hat \delta}\omega_{L{} \hat \gamma}
 R_{\alpha \hat \delta \tau}{}^{\beta} P^{\tau \hat \gamma} +
(\tilde s_{R}\tilde s_{L}\phi)\omega_{R \alpha}
P^{\alpha \hat \alpha}\omega_{L \hat \alpha }]
\label{A.10}
\end{eqnarray}
\begin{eqnarray}
 I^{(e 2)}_{gf} = {1 \over 2 }\int e^{\phi} \omega_{R \alpha}P^{\alpha
\hat \alpha} [(\tilde s_{L}\tilde s_{R} +
\tilde s_{R} \tilde s_{L})\omega_{L \hat \alpha}) +
(\tilde s_{R}\tilde s_{L}\phi)\omega_{L \hat \alpha}]
\nonumber
\end{eqnarray}
\begin{eqnarray}
=  \int e^{\phi}[ \lambda_{L}{}^{\alpha}
\omega_{R{} \beta}\lambda_{R}{}^{\hat \delta}\omega_{L{} \hat \gamma}
 R_{\alpha \hat \delta \hat \tau}{}^{\hat \gamma} P^{\beta \hat \tau} +
\omega_{R\alpha} P^{\alpha \hat \alpha}
(\tilde s_{L}\tilde s_{R}\phi) \omega_{L \hat \alpha}]
\label{A.11}
\end{eqnarray}
Now, instead of $ I^{(e 1)}_{gf} - I^{(e 2)}_{gf}$, let us consider
$ I^{(e 1)}_{gf} + I^{(e 2)}_{gf}$ .
$$  I^{(e 1)}_{gf} + I^{(e 2)}_{gf} = {1 \over 2 }\int e^{\phi}[((\tilde s_{R}
\tilde s_{L} +
\tilde s_{L}\tilde s_{R})\omega_{R \alpha})
P^{\alpha \hat \alpha}\omega_{L \hat \alpha} + ( \omega_{R \alpha}P^{\alpha
\hat \alpha}(\tilde s_{L}\tilde s_{R} +
\tilde s_{L}\tilde s_{R})\omega_{L \hat \alpha})]$$ so that
\begin{eqnarray}
  I^{(e 1)}_{gf} + I^{(e 2)}_{gf} =  \int e^{\phi} \lambda_{L}{}^{\alpha}
\omega_{R{} \beta}\lambda_{R}{}^{\hat \delta}\omega_{L{} \hat \gamma}
 [R_{\alpha \hat \delta \tau}{}^{\beta} P^{\tau \hat \gamma} + R_{\alpha
\hat \delta \hat \sigma}{}^{\hat \gamma}P^{\beta \hat \sigma}]
\label{A.12}
\end{eqnarray}
{}From this equation one can compute $\Psi_{R/L \beta \hat
\delta}{}^{\gamma
\hat \gamma}$ following the same argument used before to
compute $\Xi_{R/L \beta \hat \delta}{}^{\gamma \hat \gamma}$.
Comparing the right hand sides of (\ref{A.7}) and (\ref{A.12}) one
finally concludes that
$$ \Psi_{R/L \beta \hat \delta}{}^{\gamma \hat \gamma}= -
\Xi_{R/L \beta \hat \delta}{}^{\gamma \hat \gamma}$$
so that
\begin{eqnarray}
I^{(d)}_{gf} + I^{(e)}_{gf} = -{1 \over 2} \int e^{\phi}  \lambda_{L}{}^{\beta}
\omega_{R{} \gamma}\lambda_{R}{}^{\hat \delta}\omega_{L{} \hat \gamma}
(C_{ \beta \hat \delta}{}^{\gamma \hat \gamma} +
\tilde C_{ \beta \hat \delta}{}^{\gamma \hat \gamma})
\label{A.13}
\end{eqnarray}
Moreover, from (\ref{A.10}), (\ref{A.11}) and taking into account
(\ref{2.23}) it follows that
\begin{eqnarray}
\tilde C_{ \beta \hat \delta}{}^{\gamma \hat \gamma} = e^{\phi}
 \Pi_{\beta \tau}{}^{\gamma \sigma} [R_{\sigma \hat \sigma \rho}{}^{\tau}
P^{\rho \hat \tau} - R_{\sigma \hat \sigma \hat \rho}{}^{\hat \tau}
P^{\tau \hat \rho} + (\Gamma^{c} P \Gamma_{c})_{\sigma \hat \sigma}
P^{\tau \hat \tau}]
\Pi_{ \hat \tau \hat \delta}{}^{\hat \sigma \hat \gamma}
\label{A.14}
\end{eqnarray}



\begin{thebibliography}{99}


\bibitem{Ber1}
N. Berkovits, {JHEP \ {\bf 0004} (2000) 018, hep-th/0001035.}

\bibitem{Ber2}
N. Berkovits, {JHEP \ {\bf 0409} (2004) 047, hep-th/0406055.}

\bibitem{Ber3}
N. Berkovits and B.C. Vallilo, {JHEP \ {\bf 0007} (2000) 015,
hep-th/0004171.}

\bibitem{Ber4}
N. Berkovits, {JHEP \ {\bf 0510} (2005) 089, hep-th/0509120.}

\bibitem{Ber5}
N. Berkovits and N. Nekrasov  {JHEP \ {\bf 0612} (2006) 029, hep-th/0609012.}

\bibitem{Ber6}
N. Berkovits, {JHEP \ {\bf 0503} (2005) 041, hep-th/0406055.},
{JHEP \ {\bf 0909} (2009) 051, hep-th/08125074.}

\bibitem{BH}
N. Berkovits and P. Howe, {Nucl. Phys. \ {\bf B635} (2002) 75, hep-th/0112160.}

\bibitem{Oda1}
I. Oda and M. Tonin, {Phys. Lett. \ {\bf B520} (2001) 398, hep-th/0109051.}

\bibitem{Cha2}
O. Chandia and M. Tonin {JHEP \ {\bf 0709} (2007) 016, hep-th/07070654.}

\bibitem{Cha1}
O. Chandia, {JHEP \ {\bf 0607} (2006) 019, hep-th/0604115.}

\bibitem{Gutt}
S. Guttenberg, { hep-th/08074968}.

\bibitem{tor1}
R. D'Auria, P. Fre', P. A. Grassi and M. Trigiante, { JHEP \ {\bf 0807}
(2008) 059, hep-th/08031703}

\bibitem{dima}
J. Gomis, D. Sorokin and L. Wulff,  JHEP \ {\bf 0908} (2009)  60,
[arXiv:0811.1566 [hep-th]].

\bibitem {Arutyunov:2008if}
G.~Arutyunov and S.~Frolov,
{JHEP {\bf 0809} (2008) 129,
 hep-th/08064940.}

\bibitem{Stef}
B.~J.~Stefanski,  {Nucl.\ Phys.\  B {\bf 808}, 8087 (2009)
hep-th/08064948 .}


\bibitem{tor2}
R. D'Auria, P. Fre', P. A. Grassi and M. Trigiante, {Phys.Rev.\ {\bf D 79},
(2009)  086001 } hep-th 08081282

\bibitem{Bonel}
 G.~Bonelli, P.~A.~Grassi and H.~Safaai,
 {JHEP {\bf 0810}, 085 (2008),
 hep-th/0808.1051.}

\bibitem{Stora}
R. Stora, {Cargese Lectures, Sept.1-15, 1983 , Nato ADV.Study.Ser.Phys. 
\ {bf 115},1 (1984).} 

\bibitem {Bon1}
L. Bonora and M. Tonin { Phys.Lett.\ {\bf B98} (1981) 48 }

\bibitem {Drag}
N. Dragon, {Z.Phys.  \ {\bf C2} (1979) 29}

\bibitem{Howe}
  P.~S.~Howe and P.~C.~West,
  {Nucl.\ Phys.\  B {\bf 238}, 181 (1984).}

\bibitem{Carr}
  J.~L.~Carr, S.~J.~J.~Gates and R.~N.~Oerter,
{ Phys.\ Lett.\  B {\bf 189}, 68 (1987).}

\bibitem{Zum}
B. Zumino, {  Lectures at the Les Houches School Aug 8 - Sept 2, 1983 }

\bibitem{Bon2}
L. Bonora and P. Cotta-Ramusino,{Commun.Math.Phys. \ {\bf 87} (1983) 589}

\bibitem{MSZ}
J. Manes, R. Stora and B. Zumino {Commun. Math. Phys. \ {\bf 102} (1985),157}

\bibitem{Ansel}
  D.~Anselmi and P.~Fre,
{  Nucl.\ Phys.\  B {\bf 392} (1993) 401 arXiv:hep-th/9208029.}

\bibitem{Baul}
  L.~Baulieu, M.~P.~Bellon and R.~Grimm,
 { Nucl.\ Phys.\  B {\bf 294}, 279 (1987).}

\bibitem{Super}
  D.~P.~Sorokin, V.~I.~Tkach and D.~V.~Volkov,
 { Mod.\ Phys.\ Lett.\  A {\bf 4}, 901 (1989),}
M. Tonin,{Phys.Lett. \ {\bf 266} (1991) 312,}
N. Berkovits, {Nucl.Phys. \  B  {\bf 350}(1991) 193}
F. Delduc,A. Galperin, P. Howe,E. Sokatchev, {Phys. Rev. \ D {\bf 47} (1993)
578}
I.~A.~Bandos, D.~P.~Sorokin, M.~Tonin, P.~Pasti and D.~V.~Volkov,
{Nucl.\ Phys.\  B {\bf 446}, 79 (1995)  hep-th/9501113,}
  P.~S.~Howe and E.~Sezgin, {Phys.\ Lett.\  B {\bf 390}, 133 (1997)
  hep-th/9607227,}
D.~P.~Sorokin, {Phys.\ Rept.\  {\bf 329}, 1 (2000)  hep-th/9906142.}

\bibitem{Y-form1}
M.~Matone, L.~Mazzucato, I.~Oda, D.~Sorokin and M.~Tonin,
 { Nucl.\ Phys.\  B {\bf 639}, 182 (2002),
  hep-th/0206104.}

\bibitem {Ber7}
N. Berkovits, { JHEP \ {\bf 0209} (2002) 051, hep-th/ 0201151}

\bibitem{Fre2}
  P.~Fr\'e and P.~A.~Grassi,
{  Nucl.\ Phys.\  B {\bf 763} (2007) 1
  arXiv:hep-th/0606171.}

\bibitem{Y-form2}
I.~Oda and M.~Tonin,
 { Phys.\ Lett.\  B {\bf 606}, 218 (2005),
 :hep-th/0409052.}

\bibitem{Y-form3}
I.~Oda and M.~Tonin,
  { Nucl.\ Phys.\  B {\bf 727}, 176 (2005),
  hep-th/0505277.}

\bibitem{Y-form4}
I.~Oda and M.~Tonin,
 {  Nucl. Phys.\  B {\bf 779}, 63 (2007),
  hep-th/07041219}

\bibitem{Sken}
J. Hoogeven and K. Skenderis, {JHEP \ {\bf 0711} (2007) 081
hep-th/ 07102598}

\end{thebibliography}
\end{document}